\documentclass[aps,prx,10pt,twocolumn,superscriptaddress,notitlepage,floatfix]{revtex4-2}

\usepackage[utf8]{inputenc}
\usepackage{graphicx}
\graphicspath{{../}}

\pdfoutput=1 

\usepackage[ruled,vlined]{algorithm2e}

\usepackage[colorlinks=true, hyperindex, breaklinks, linkcolor=blue, urlcolor=blue, citecolor=blue]{hyperref} 

\usepackage{dsfont}
\usepackage{amsmath}
\usepackage{amssymb,bm}
\usepackage{amsthm}
\usepackage{array, makecell}
\usepackage{boldline}
\usepackage[capitalise]{cleveref}
\usepackage{graphicx}
\usepackage[caption=false]{subfig}
\usepackage[normalem]{ulem}
\usepackage{cancel}

\begin{document}
\clearpage

\title{Preserving phase coherence and linearity in cat qubits with exponential bit-flip suppression}

\author{Harald Putterman} 
\thanks{These authors contributed equally}
\email{putterma@amazon.com}
\author{Kyungjoo Noh} 
\altaffiliation[]{These authors contributed equally}
\author{Rishi N. Patel}
\author{Gregory A. Peairs} 
\author{Gregory S. MacCabe} 
\author{Menyoung Lee} 
\author{Shahriar Aghaeimeibodi} 
\author{Connor T. Hann} 
\author{Ignace Jarrige} 
\author{Guillaume Marcaud} 
\author{Yuan He}
\author{Hesam Moradinejad} 
\author{John Clai Owens} 
\author{Thomas Scaffidi} 
\altaffiliation[Current affiliation: ]{Department of Physics and Astronomy, University of California, Irvine, California 92697, USA.}
\author{Patricio Arrangoiz-Arriola} 
\author{Joe Iverson}
\author{Harry Levine} 
\affiliation{AWS Center for Quantum Computing, Pasadena, CA 91125, USA}
\author{Fernando G.S.L. Brandão} 
\affiliation{AWS Center for Quantum Computing, Pasadena, CA 91125, USA}
\affiliation{IQIM, California Institute of Technology, Pasadena, CA 91125, USA}
\author{Matthew H. Matheny} 
\affiliation{AWS Center for Quantum Computing, Pasadena, CA 91125, USA}
\author{Oskar Painter} 
\email{ojp@amazon.com}
\affiliation{AWS Center for Quantum Computing, Pasadena, CA 91125, USA}
\affiliation{IQIM, California Institute of Technology, Pasadena, CA 91125, USA}
\affiliation{Thomas J. Watson, Sr., Laboratory of Applied Physics,
California Institute of Technology, Pasadena, California 91125, USA}

\date{\today}

\begin{abstract}

Cat qubits, a type of bosonic qubit encoded in a harmonic oscillator, can exhibit an exponential noise bias against bit-flip errors with increasing mean photon number.  Here, we focus on cat qubits stabilized by two-photon dissipation, where pairs of photons are added and removed from a harmonic oscillator by an auxiliary, lossy buffer mode. This process requires a large loss rate and strong nonlinearities of the buffer mode that must not degrade the coherence and linearity of the oscillator. In this work, we show how to overcome this challenge by coloring the loss environment of the buffer mode with a multi-pole filter and optimizing the circuit to take into account additional inductances in the buffer mode. Using these techniques, we achieve near-ideal enhancement of cat-qubit bit-flip times with increasing photon number, reaching over $0.1$~seconds with a mean photon number of only $4$. Concurrently, our cat qubit remains highly phase coherent, with phase-flip times corresponding to an effective lifetime of $T_{1,\text{eff}} \simeq 70$~$\mu$s, comparable with the bare oscillator lifetime. We achieve this performance even in the presence of an ancilla transmon, used for reading out the cat qubit states, by engineering a tunable oscillator-ancilla dispersive coupling.  Furthermore, the low nonlinearity of the harmonic oscillator mode allows us to perform pulsed cat-qubit stabilization, an important control primitive, where the stabilization can remain off for a significant fraction (e.g., two thirds) of a $3~\mathrm{\mu s}$ cycle without degrading bit-flip times. These advances are important for the realization of scalable error-correction with cat qubits, where large noise bias and low phase-flip error rate enable the use of hardware-efficient outer error-correcting codes.

\end{abstract}

\maketitle

\section{Introduction}

The expected system size required for realizing fault-tolerant quantum computers far exceeds the scale of current experiments~\cite{OGorman2017,Kivlichan2020,Gidney2021}. One approach to reduce the resource overhead for fault tolerance is to use qubits which have a biased noise structure (e.g., bit-flip error rates are much smaller than the phase-flip error rates).  This noise bias property allows the quantum error correcting code to focus predominantly on correcting only one type of error, for example phase-flip errors, leading to reduced overhead for error correction~\cite{Cochrane1999Macroscopically,Aliferis2008,Tuckett2018_ultrahigh,Guillaud2019Repetition,Guillaud2021_error,Darmawan2021,BonillaAtaides2021_XZZX,chamberland2020building,Regent2023highperformance,Gouzien2023,Ruiz2024}.  

\begin{figure}[b!]
    \centering
    \includegraphics[width=\columnwidth]{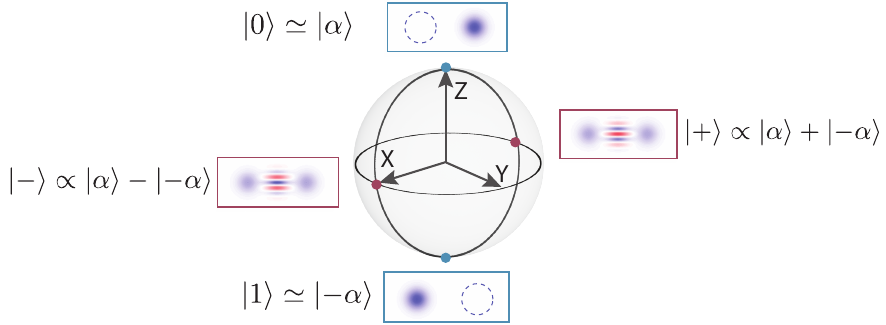}
    \caption{\textbf{Bloch sphere of a cat qubit.} Wigner functions of the computational and complementary basis states are represented next to the corresponding points on the Bloch sphere. 
    }
    \label{fig:cat_visual}
\end{figure}

Cat qubits~\cite{Cochrane1999Macroscopically,Jeong2002Efficient} are a specific realization of biased-noise qubits. In cat qubits, the quantum information is encoded into the infinite-dimensional Hilbert space of a quantum harmonic oscillator.  The  computational basis states of a cat qubit are approximately equivalent to the coherent states, $|0\rangle\simeq |+\alpha\rangle$ and $|1\rangle \simeq |-\alpha\rangle$.  The complementary basis states are exactly given by the even- and  odd-parity cat states, i.e., $|\pm\rangle = |C^{\pm}_{\alpha}\rangle \propto |+\alpha\rangle \pm |-\alpha\rangle$ (see \cref{fig:cat_visual}). Cat qubits can have exponentially suppressed bit-flip ($X$) error rates with increasing mean photon number $|\alpha|^2$~\cite{Lescanne2020,Berdou2023,MarquetAutoparametric2024,Reglade2024}. This is due to the phase-space separation between the two computational basis states which gets larger as $|\alpha|$ increases.  In contrast, the phase-flip ($Z$) error rates of a cat qubit increase only linearly with the mean photon number $|\alpha|^2$.

To realize this exponential noise bias against bit-flip errors, it is necessary to stabilize the oscillator mode in the cat-qubit state manifold.  In this work, we focus on cat qubits stabilized with two-photon dissipation, so called ``dissipative cat qubits''~\cite{mirrahimi2014,leghtas2015,touzard2018}.  Dissipative cat qubits are implemented by engineering a nonlinear coupling between an oscillator mode (also referred to as a storage mode) and a lossy buffer mode.  The nonlinear coupling mediates an exchange of two photons from the storage to one photon in the buffer mode via a three-wave mixing (3WM) nonlinearity. By adding a linear buffer drive and adiabatically eliminating the lossy buffer mode, we realize a two-photon dissipation described by the master equation
\begin{align}
    \frac{d\hat{\rho}}{dt}=\kappa_2 D[\hat{a}^2-\alpha^2]\hat{\rho} , \label{eq:ideal_two_photon_dissipation_master_equation}
\end{align}
which stabilizes the oscillator mode $\hat{a}$ in a two-dimensional steady-state subspace spanned by the coherent states $|\pm\alpha\rangle$~\cite{mirrahimi2014}.

For cat qubits to enable reduced error correction overhead, it is essential to achieve an exponential noise bias without sacrificing the cat qubit's phase coherence. Specifically, a large noise bias must be achieved at small values of $|\alpha|^{2}$ where the phase-flip error rate remains low enough to be correctable by an outer error-correcting code such as a repetition code~\cite{Guillaud2019Repetition,chamberland2020building,Guillaud2021_error,Regent2023highperformance}. Achieving this requires preserving the coherence and linearity of the storage mode under the two-photon dissipation. This is challenging due to the strong nonlinear 3WM coupling of the storage mode to the highly lossy buffer mode used for dissipative stabilization. Additionally, in some architectures, interactions needed to characterize the storage mode can also create undesired nonlinearities.  While there has been encouraging recent progress addressing these challenges, prior experiments have so far demonstrated either high cat phase coherence~\cite{touzard2018} or exponential noise bias~\cite{Lescanne2020,Reglade2024,MarquetAutoparametric2024}, but not both simultaneously. 

In this paper, we present a cat qubit system where the bit-flip error rates are exponentially suppressed and the cat qubit remains highly phase coherent. In particular, we achieve high effective storage-mode lifetimes of $T_1\simeq  70\mu s$ by using a multipole metamaterial filter to protect the storage from the strong buffer mode loss channel. Simultaneously, our device employs two novel strategies to preserve the storage linearity.  First, a tunable coupler with high on-off ratio dispersive coupling enables characterization of the storage mode with an ancillary transmon without introducing deleterious nonlinearities. Second, we engineer an advantageous cancellation of buffer-induced nonlinearities by carefully accounting for serial inductances present in our buffer mode circuit. Together, the storage mode's high coherence and low nonlinearity enable us to demonstrate large noise bias at low photon numbers, as is crucial for hardware-efficient error correction. We measure exponential bit-flip suppression, with near-ideal scaling~\cite{mirrahimi2014}, up to $0.1~\text{s}$ bit-flip times at $|\alpha|^{2}=4$. Furthermore, when pulsing the two-photon dissipation on and off with $3~\mathrm{\mu s}$ cycles, we are able to achieve bit-flip times comparable to the continuous case despite the fact that the stabilization is on only one third of the time during a cycle.  These advances are key enablers for performing more complex operations on cat qubits such as  multi-qubit gates and error-correction  circuits. 

This paper is organized as follows. After providing an overview of our device in \cref{sec:device_overview}, we present the key technical advances necessary for realizing a high-coherence and low-nonlinearity storage mode. First, in \cref{sec:tunable_dispersive_coupling}, we introduce a tunable dispersive coupling with a high on-off ratio for reading out the storage mode state, where the tunability protects the storage mode from errors due to the readout components. Next, in \cref{sec:storage_mode_coherence} we address the tradeoff between the required strong buffer loss and storage coherence by coloring the buffer environment with a multi-pole metamaterial filter. Finally, in \cref{sec:storage_linearity} we discuss the importance of accounting for serial inductances within the buffer to realize storage modes with low nonlinearity. Equipped with these technical advances, we then benchmark the performance of our cat qubit. We demonstrate the  cat qubit's exponential noise bias in \cref{sec:two_photon_dissipation_performance} and the persistence of that noise bias under pulsed stabilization in \cref{sec:pulsed_stabilization}.

\section{Device overview}
\label{sec:device_overview}

\begin{figure}[t!]
    \centering
    \includegraphics[width=\columnwidth]{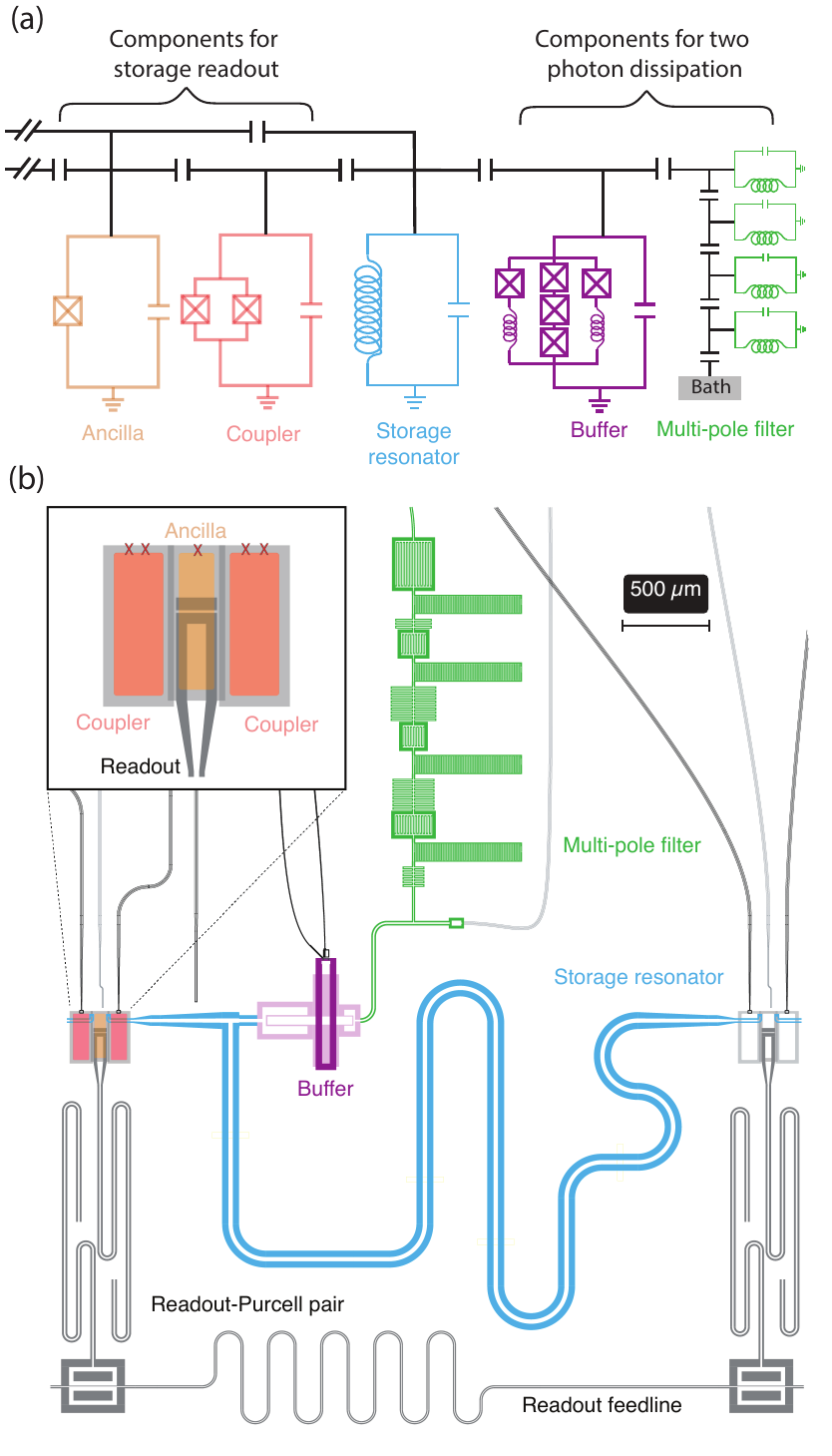}
    \caption{\textbf{Device overview.} (a) Circuit representation of the key components of a dissipative cat qubit with ancilla transmon.  The storage mode is read out by an ancilla transmon using a tunable-transmon coupler.  The storage mode is also coupled to a buffer mode.  The buffer is coupled to a 4-pole metamaterial filter consisting of an array of coupled LC oscillators. The buffer and filter modes are used for implementing two-photon dissipation to stabilize a cat qubit in the storage mode. (b) Physical layout of the device.  The storage mode is a $\lambda/2$ resonator.  The metamaterial filter is implemented as a series of lumped LC oscillators. }   
    \label{fig:device_overview}
\end{figure}

We consider a system containing a storage mode $\hat{a}$, a buffer mode $\hat{b}$, and components for reading out the state of the storage mode as shown in \cref{fig:device_overview}. The central element of our system is the storage mode (blue in \cref{fig:device_overview}), where the cat qubit states are encoded. The storage mode is implemented as a $\lambda/2$ resonator and has frequency $\omega_s/2\pi=5.35~\text{GHz}$. 
 
The buffer mode (purple in \cref{fig:device_overview}) is implemented using a variant of the Asymmetrically-Threaded SQUID (ATS) element~\cite{Lescanne2020}, consisting of a middle junction array and two side junctions. Two noteworthy aspects of our buffer design are the shallowness of the middle junction array (in the sense that it only has three junctions in the array), and the fact that each side junction has a small but non-negligible serial inductance. The significance of these design choices are discussed in detail in \cref{sec:storage_linearity}.  The buffer has two flux lines controlling the fluxes through the two loops of the ATS, one symmetrically ($\varphi_\Sigma$) the other differentially ($\varphi_\Delta$).  It is also coupled to a lossy environment through an output line colored by a 4-pole metamaterial bandpass filter~\cite{ferreira2020collapse, Mirhosseini2018} (green in \cref{fig:device_overview}), with the passband ranging from $\sim 2.6~\text{GHz}$ to $\sim 3.6~\text{GHz}$.  At the chosen operating point, the frequency and the loss rate of the buffer mode are given by $\omega_b/2\pi=3.01~\text{GHz}$ and $\kappa_b/2\pi=10.7~\text{MHz}$, respectively. 

The storage mode is read out by an ancilla transmon (yellow in \cref{fig:device_overview}) with a dedicated readout resonator. The storage mode is dispersively coupled to the ancilla transmon through a tunable-transmon coupler (red in \cref{fig:device_overview}). The frequency of the coupler can be tuned by varying the external flux of the coupler's SQUID loop. This tunability allows us to selectively turn on the dispersive coupling between the storage mode and the ancilla qubit only when the storage mode needs to be read out (see \cref{sec:tunable_dispersive_coupling}).

The device consists of two silicon chips which are flip-chip bonded together~\cite{Foxen2018, Das2018}. The bottom chip hosts the linear elements including the storage mode and uses a Tantalum superconducting thin-film layer~\cite{Place2021}. The top chip has the non-linear elements containing Josephson junctions such as the buffer mode, the ancilla transmon, and the tunable-transmon coupler, and uses a series of Aluminum superconducting thin-film layers. The schematic of \cref{fig:device_overview}(a) shows the circuit diagram of our device, with the circuit layout shown in \cref{fig:device_overview}(b). As indicated in the circuit layout, the cat qubit device studied here is part of a larger integrated circuit involving an array of cat qubits, with each storage mode coupled two ancillas qubits, one on the left and one on the right.

\section{Tunable dispersive coupling for reading out a storage mode}
\label{sec:tunable_dispersive_coupling}

\begin{figure}
    \centering
    \includegraphics[width=\columnwidth]{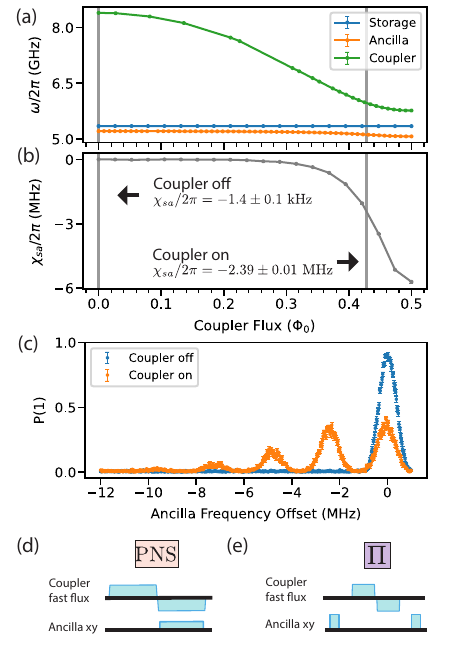}
    \caption{\textbf{Tunable dispersive coupling for reading out a storage mode.}  (a) Frequencies of the coupler, ancilla, and storage mode as a function of the external flux of the coupler.  (b) Storage-ancilla dispersive coupling as a function of the coupler flux measured using a storage conditional phase experiment.  When the tunable coupler is pulsed to the on-position with $\Phi_{x,c} = 0.43\Phi_{0}$, the storage and ancilla transmon develop a dispersive coupling $|\chi_{sa}|/2\pi > 2~\text{MHz}$.  Using an alternative measurement method (based on Ramsey experiments on the ancilla) we measure the $\chi_{sa}/2\pi = -1.4\pm 1 ~\text{kHz}$ in the off-position with $\Phi_{x,c} = 0$. (c) Number-splitting spectroscopy on the ancilla transmon with the coupler either in the on- or off-position and the storage mode in a coherent state. (d) Photon-number-selective (PNS) pulse sequence on the ancilla used in \cref{fig:tunable_coupler}(c) to measure the storage mode. (e) Storage parity mapping ($\Pi$) pulse sequence used in, for example, \cref{fig:storage_coherence}(c). Both storage mode measurement sequences in (d) and (e) use net-zero waveforms for the tunable coupler flux pulse. }   
    \label{fig:tunable_coupler}
\end{figure}

A commonly used strategy for reading out quantum states of a linear oscillator mode is to couple it dispersively to an ancillary qubit. Through the dispersive coupling, the ancilla qubit experiences frequency shifts depending on the number of excitations in the oscillator mode. Such frequency shifts can then be used to characterize the state of the oscillator mode~\cite{lutterbach1997,Schuster2007,Leghtas2013,sun2014}. However, this same dispersive coupling can also introduce unwanted frequency shifts of the oscillator mode when the number of excitations in the ancilla qubit changes due to, for instance, decay or heating. Such unwanted frequency shifts have been one of the main performance bottlenecks for various error correction protocols with oscillator modes~\cite{ofek2016,Lescanne2020,chou2023}. In this section, we provide a solution to this problem by engineering tunable dispersive coupling between a storage mode and an ancilla transmon via a tunable-transmon coupler.

Here, we focus on a subsystem comprising a storage mode, a tunable-transmon coupler, and a fixed-frequency ancilla transmon which are capacitively coupled (see \cref{app:tunable_dispersive_coupling_details}). The effective Hamiltonian of this subsystem takes the form  
\begin{align}
    \hat{H} &= \omega_{s}\hat{a}^{\dagger}\hat{a} + \omega_{c}|e_{c}\rangle\langle e_{c}| + \omega_{a}|e_{a}\rangle\langle e_{a}| + \chi_{sa}\hat{a}^{\dagger}\hat{a}|e_{a}\rangle\langle e_{a}|,  \label{eq:storage_coupler_ancilla_effective_hamiltonian}
\end{align}
in the dressed eigenbasis where all the parameters $\omega_{s}, \omega_{c}, \omega_{a}$ and $\chi_{sa}$ depend on the external flux of the coupler, $\Phi_{x,c}$. Here, $|e_{c}\rangle$ and $|e_{a}\rangle$ refer to the first excited state of the coupler and the ancilla, respectively. Thus, $\omega_{s}$, $\omega_{c}$, and $\omega_{a}$ are the frequencies of the storage, coupler, and ancilla in the lowest energy manifold consisting of the ground state and the first excited state. Most importantly, $\chi_{sa}$ is the strength of the dispersive coupling between the storage and the ancilla, i.e., the frequency shift of the storage mode when the ancilla is excited from the ground state to its first excited state $|e_{a}\rangle$.  

The external flux of the coupler, $\Phi_{x,c}$, tunes the frequency of the coupler mode. In \cref{fig:tunable_coupler}(a), we show the frequencies of the storage, ancilla, and coupler as a function of $\Phi_{x,c}$. The frequency of the coupler tunes from $\omega_{c,\text{max}} / 2\pi = 8.40~\text{GHz}$ to $\omega_{c,\text{min}} / 2\pi = 5.77~\text{GHz}$ as the coupler external flux is varied from $0$ to the half flux quantum $\Phi_{0}/2$. Note that the storage and ancilla frequencies are also lowered as the coupler approaches its minimum frequency. This is due to the increased mode hybridization with the coupler and the associated repulsion between the energy levels. 

We select the idle (off-) position to be when the coupler is at its maximum frequency with $\Phi_{x,c}=0$. To resolve the small dispersive coupling between the storage and ancilla with the coupler in the off-position, we perform Ramsey measurements of the ancilla transmon with a variable mean photon number coherent state in the storage mode (see \cref{app:off_position_chi}). In particular, a mean photon number up to $20$ is used to amplify the impact of the dispersive coupling on the ancilla. With this method, we find that the strength of the dispersive coupling is only $\chi_{sa} / 2\pi = -1.4 \pm 0.1 ~\text{kHz}$ in the off-position. 

We can dynamically turn on the dispersive coupling by applying a flux pulse on the tunable coupler (see ``coupler fast flux'' in \cref{fig:tunable_coupler}(d)). By pulsing the coupler to lower frequencies, we increase the hybridization between modes and consequently increase the dispersive coupling between the storage and ancilla. In \cref{fig:tunable_coupler}(b), we characterize the dispersive coupling at each flux amplitude by measuring the difference in phase accumulation rate on the storage mode conditioned on the state of the ancilla (see \cref{app:storage_conditional_phase} for more details). The maximum dispersive coupling is achieved with the coupler at its minimum frequency, where $|\chi_{sa}|/2\pi >5\ \mathrm{MHz}$.

In practice, we do not use the coupler's minimum frequency position ($\Phi_{x,c} = \Phi_{0} / 2$) as the dispersive nature of the interaction breaks down in the large storage photon number regime due to strong hybridization (see \cref{app:tunable_dispersive_coupling_details}).  Instead, we select an on-position to be $\Phi_{x,c} = 0.43\Phi_{0}$ where we observe a dispersive coupling of $\chi_{sa}/2\pi=-2.39 \pm 0.01~\text{MHz}$. Relative to the off-position $\chi_{sa}/2\pi$ of $-1.4~\text{kHz}$, this consitutes a dynamic range of $>1000$ for the storage-ancilla dispersive coupling. The high on-off ratio is made possible by carefully optimizing the direct capacitive coupling between the storage and ancilla so that the net dispersive interaction between them is minimized around the idle off-position of the tunable coupler~\cite{Yan2018Tunable, Sung2021CZ}.

The tunability of the dispersive interaction is further demonstrated in \cref{fig:tunable_coupler}(c), through number-splitting spectroscopy of the ancilla transmon. See \cref{fig:tunable_coupler}(d). Here, the storage mode is driven into a coherent state and a weak $\pi$-pulse (length $1.4~\mathrm{\mu s}$) is applied to the ancilla transmon. By varying the drive frequency of the ancilla $\pi$-pulse, we measure the spectrum of the ancilla transmon.  When the coupler is in the off-position (i.e., no coupler flux pulse applied), we cannot resolve the splitting of the ancilla frequency, as expected and desired. When the coupler is pulsed to the on-position, we see distinct peaks in the ancilla spectrum, corresponding to different Fock states of the storage mode. The spacing between the neighboring peaks is consistent with the dispersive shift, $\chi_{sa}$, measured in \cref{fig:tunable_coupler}(b).

\begin{figure*}[t!]
    \centering
    \includegraphics[width=\textwidth]{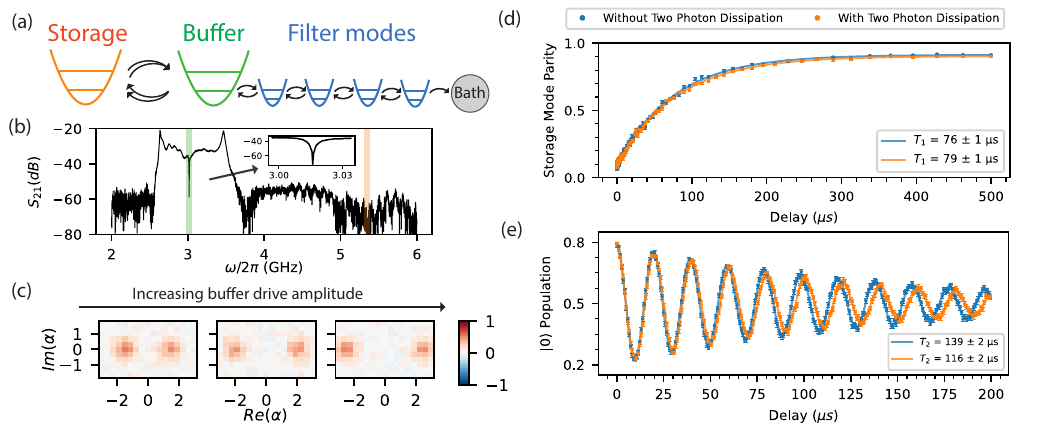}
    \caption{\textbf{Engineering two-photon dissipation while preserving storage mode coherence.} (a) Schematic diagram of a storage mode which is coupled to a lossy buffer mode by a three-wave mixing process exchanging two photons from the storage mode with one photon in the buffer mode. The strong loss on the buffer mode is engineered by coupling the buffer mode to a lossy transmission line through a 4-pole metamaterial filter. (b) The transmission spectrum through the 4-pole metamaterial filter.  The metamaterial filter colors the environment seen by the buffer mode.  The buffer frequency (green) lies within the passband resulting in strong damping of the buffer mode.  The storage-mode frequency (orange) lies outside of the filter passband resulting in suppression of storage mode loss to the filtered environment. 
    Note that this $S_{21}$ data is convolved with the uncalibrated response of a $2.5-4$~GHz isolator so the filter spectrum may be different than that reported outside the $2.5-4$~GHz range, and thus at the storage mode frequency. Nonetheless the storage loss is strongly filtered, with electromagnetic simulations predicting a storage loss rate due to the filter of $\sim 0.2~\mathrm{kHz}$. (c) Wigner tomograms of the steady state cat qubit manifolds of the two-photon dissipation with increasing buffer drive amplitude.  The coherent state amplitude of a cat qubit in the storage mode increases as the amplitude of the linear buffer drive increases. (d) Measured $T_{1}$ of the storage mode with and without the pure two-photon dissipation ($\kappa_{2}\mathcal{D}[\hat{a}^{2}]$, i.e., without the buffer drive) being applied during delay. (e) Measured $T_{2}$ of the storage mode with and without the pure two-photon dissipation being applied during delay.    }
    \label{fig:storage_coherence}
\end{figure*}

As shown in \cref{fig:tunable_coupler}(e), the tunable dispersive coupling also enables parity measurement of the storage mode by combining a Ramsey measurement on the ancilla with a flux pulse ($\Phi_{x,c}/\Phi_{0} = 0.38$; see \cref{app:device_parameters}) on the coupler. In the parity measurement, one needs to apply unselective $\pi/2$ pulses on the ancilla, as well as a conditional (on the ancilla being excited) $\pi$-rotation of the storage mode through the storage-ancilla dispersive coupling.  Due to the large on-off ratio, the unselective $\pi/2$ pulses are easy to achieve since they are applied when the storage-ancilla dispersive coupling is turned off and the resonant frequency of the ancilla is unperturbed by the storage-mode state. The parity measurement allows us to perform Wigner tomography~\cite{lutterbach1997} of the storage mode (see, e.g., \cref{fig:storage_coherence}(c)). Further advantages of the tunable dispersive coupling in the specific context of a cat qubit are demonstrated in \cref{sec:two_photon_dissipation_performance}. 

\section{Engineering two-photon dissipation while preserving storage mode coherence}
\label{sec:storage_mode_coherence}

Equipped with the ability to read out the storage mode state, we now describe the engineering of two-photon dissipation for stabilizing a cat qubit. Two-photon dissipation can be realized by engineering a 3WM Hamiltonian $\hat{H}_{\text{3WM}} = g_{2}\hat{a}^{2}\hat{b}^{\dagger} + \text{H.c.}$, a linear drive $\hat{H}_{d} = -g_{2}\alpha^{2}\hat{b}^{\dagger}$ + \text{H.c.}, and a loss $\kappa_{b}\mathcal{D}[\hat{b}]$ on the buffer mode. After adiabatically eliminating the buffer mode,  the dynamics of this system obey \cref{eq:ideal_two_photon_dissipation_master_equation} with $\kappa_{2} \simeq 4g_{2}^{2} / \kappa_{b}$. Here, the key technical challenge is the implementation of the 3WM Hamiltonian which requires a strong flux pump on the buffer mode and large buffer loss rate. The strong driven-dissipative processes on the buffer mode should not degrade the coherence of the storage mode.  

In order to protect the storage mode from the strong loss channel of the buffer mode, we color the loss environment of the buffer mode with a multi-pole metamaterial filter~\cite{Mirhosseini2018,ferreira2020collapse, chamberland2020building,putterman2022}. Unlike commonly used Purcell filters composed of only one resonator mode, our filter consists of four lumped-element resonator modes which are capacitively coupled. It can also be understood as a series of coupled oscillators as illustrated in \cref{fig:storage_coherence}(a).

The use of a 4-pole filter allows us to realize a passband with a bandwidth of about $1~\text{GHz}$ with sharp band edges (see \cref{fig:storage_coherence}(b)). The broad bandwidth makes the system resilient against mistargeting of the buffer frequency. Frequency offsets on the order of $100~\text{MHz}$ can be tolerated since the buffer loss rate is maintained at around $10~\text{MHz}$ throughout much of the passband. The storage mode, however, is far outside of the filter passband and therefore is protected against loss through the filter. From electromagnetic simulations, we estimate that the induced loss rate on the storage mode due to the buffer loss channel is $\sim 0.2~\text{kHz}$, over four orders of magnitude smaller than the buffer loss rate of $\kappa_b/2\pi=10.7~\text{MHz}$. This corresponds to a $T_{1}$ ceiling of $ > 750$~$\mu$s on the storage mode. 

In \cref{fig:storage_coherence}(c), we demonstrate stabilization of the cat qubit manifold by integrating the 3WM pump, linear drive, and loss on the buffer mode. In particular, we turn on the 3WM interaction with strength $g_{2}/2\pi = 578\pm 4~\text{kHz}$ (corresponding to $\kappa_{2}/2\pi = 124\pm 2~\text{kHz}$; see \cref{app:g2_fitting}) 
and a linear buffer drive with a variable amplitude.  As the strength of the linear buffer drive increases, the distance between the two steady states (i.e., the ``size'') of the cat qubit increases. When the buffer drive is turned off while the 3WM pump is kept on, a pure two-photon dissipation $\kappa_{2}\mathcal{D}[\hat{a}^{2}]$ is implemented which confines the storage mode into the $|\hat{n}=0\rangle/|\hat{n}=1\rangle$ Fock-state manifold. Note that we use the term pure two-photon dissipation to refer to the dynamics due to $\kappa_{2}\mathcal{D}[\hat{a}^{2}]$ with $|\alpha|=0$, to distinguish it from the case with $|\alpha|>0$.

We study the lifetime ($T_1$) and coherence time ($T_2$) of the storage mode in this integrated system using the following procedure. We prepare the state  $(|\hat{n}=0\rangle + |\hat{n}=1\rangle) / \sqrt{2}$ (up to state-preparation errors and phase) in the confined Fock-state manifold of the storage mode by initializing a large coherent state and applying a pure two-photon dissipation to the storage mode~\cite{MarquetAutoparametric2024}. We then let the storage mode evolve, either with the pure two-photon dissipation turned off or kept on, for a variable delay.  Ideally, the pure two-photon dissipation acts trivially within the  $|\hat{n}=0\rangle$/$|\hat{n}=1\rangle$ manifold so it should not affect the storage-mode coherence. However, strong off-resonant driving of various unintended transitions can degrade the storage mode coherence.

To characterize the storage $T_{1}$ time, after the variable delay time, we measure the photon-number parity of the storage mode and extract the decay rate of the parity observable.  As shown in \cref{fig:storage_coherence}(d), the storage $T_1$ with the pure two-photon dissipation turned off is measured to be $76 \pm 1$~$\mu$s. This relatively long $T_1$ value is achieved due to: (i) the optimized deposition and etching of the tantalum superconducting thin-film used in the fabrication of the chip which hosts the storage mode coplanar waveguide resonators~\cite{Place2021, Wang2022}, and (ii) the effective protection of the storage mode by the 4-pole filter. Furthermore, we observe that the storage $T_{1}$ is not degraded when the pure two-photon dissipation is kept on during the delay. This can be attributed to careful allocation of the storage and buffer mode frequencies to avoid undesired resonances under a strong 3WM flux pump (see \cref{app:pump_loss}). 

The storage $T_{2}$ is extracted through a Ramsey-style measurement where after the preparation into $(|\hat{n}=0\rangle + |\hat{n}=1\rangle) / \sqrt{2}$ and variable delay, we apply a storage displacement, and a final measurement of the storage mode's vacuum state probability using a photon-number-selective pulse on the ancilla.  As shown in \cref{fig:storage_coherence}(e), the storage mode coherence time, $T_2$, is measured to be over $100~\mathrm{\mu s}$ with the pure two-photon dissipation turned off. With the dissipation turned on, $T_2$ is only marginally degraded. These results demonstrate a highly coherent storage mode which can host a stabilized, phase-coherent cat qubit. See \cref{app:storage_t1_t2_msmts} for more details on the $T_{1}$ and $T_{2}$ measurements.

\section{Buffer induced storage mode nonlinearities}
\label{sec:storage_linearity}

Realizing a cat qubit with an exponential noise bias and a low phase-flip error rate requires more than high storage-mode coherence times.  It also requires minimizing undesired, parasitic nonlinearities in the storage mode. In particular, we identify the buffer-induced nonlinearities (e.g., the storage self-Kerr and the storage-buffer cross-Kerr) as the most crucial parameters. We then discuss how the serial inductances in the buffer mode can be utilized to minimize these parasitic nonlinearities. 
 
In the presence of the storage-buffer cross-Kerr and storage self-Kerr, the Hamiltonian of the storage-buffer subsystem is given by
\begin{align}
    \hat{H} &= \Big{(}g_{2}(\hat{a}^{2}-\alpha^{2})\hat{b}^{\dagger} + \text{H.c.}\Big{)}  + \chi_{sb}\hat{a}^{\dagger}\hat{a}\hat{b}^{\dagger}\hat{b} + \frac{K_{s}}{2}\hat{a}^{\dagger 2}\hat{a}^{2} . 
    \label{eq:simple_two_photon_main_text}
\end{align} 
The first two terms are the resonant 3WM term and linear drive on the buffer mode, which together stabilize a cat qubit. The last two terms are the undesired storage-buffer cross-Kerr and storage self-Kerr. By adding the buffer loss $\kappa_{b}\mathcal{D}[\hat{b}]$, and adiabatically eliminating the buffer mode, the effective master equation within the storage mode is approximately given by
\begin{align}
    \frac{d\hat{\rho}}{dt}=-i[\hat{H}_{\text{eff}},\hat{\rho}]+\frac{4g_2^2}{\kappa_b}D\Big{[}\Big{(}1-\frac{2i\chi_{sb}}{\kappa_b}\hat{a}^\dagger \hat{a}\Big{)}(\hat{a}^2-\alpha^2)\Big{]}\hat{\rho}, 
\end{align}
with
\begin{align}
    \hat{H}_{\text{eff}}=-\frac{2 g_2^2\chi_{sb}}{\kappa_b^2}(\hat{a}^{\dagger 2}-\alpha^2)\hat{a}^\dagger \hat{a}(\hat{a}^2-\alpha^2) + \frac{K_{s}}{2} \hat{a}^{\dagger 2} \hat{a}^2 , 
\end{align}
in the regime $\chi_{sb} \langle \hat{a}^\dagger \hat{a}\rangle  /\kappa_b  \ll 1$ (see \cref{app:effective_model_analysis}).

\begin{figure*}
    \centering
    \includegraphics[width=\textwidth]{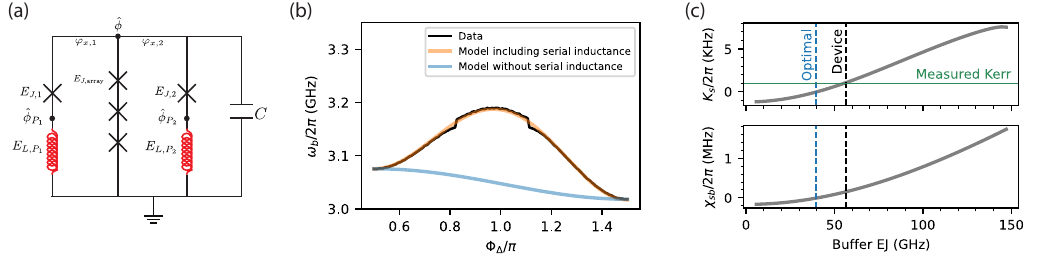}
    \caption{\textbf{Buffer-induced storage mode nonlinearities.} (a) Circuit diagram of the buffer mode, including additional serial inductances (colored in red) in series with the side junctions. In the central path we use a shallow 3-element junction array.  (b) Frequency of the buffer mode as a function of $\varphi_\Delta$ with $\varphi_\Sigma=\pi/2$.  Adding serial inductances to the model is necessary to accurately reproduce the experimental data.   (c) Numerical predictions based on a model where the underlying circuit parameters are tuned to match (b).  We let $E_J$ of the model vary and plot the predictions of storage self-Kerr (top plot) and the storage-buffer cross-Kerr (bottom plot) as a function of the average buffer side junction energy $E_J = (E_{J,1} + E_{J,2})/2$.  The value of $E_J$ extracted from fitting (b) (black dashed line) is close to the optimal value of $E_J$ (blue dashed line).  To verify the model we measure the storage self-Kerr to be $0.97\pm 0.02~\text{kHz}$ (green horizontal line), which is close to the model predicted value of $K_{s}/2\pi = 1.1~\text{kHz}$ at the independently extracted device $E_J$ value.}   
    \label{fig:serial_inductance}
\end{figure*}

The effective master equation above clearly shows the detrimental effects of the storage-buffer cross-Kerr $\chi_{sb}$. Specifically, the two-photon dissipation operator gets distorted by a factor of $(1 - 2i\chi_{sb}\hat{a}^{\dagger}\hat{a} / \kappa_{b})$, and an additional term is added to the Hamiltonian $\hat{H}_{\text{eff}}$ which acts non-trivially outside of the cat-qubit manifold. An intuitive reason for the particularly damaging effect of these undesired terms is as follows. Noise makes the storage mode deviate from the cat qubit manifold which then excites the buffer mode. This buffer excitation should only be accompanied by an ideal two-photon dissipation operator $\hat{a}^{2}-\alpha^{2}$ on the storage mode. However, the buffer excitation additionally dephases the storage mode due to the storage-buffer cross-Kerr. This extra dephasing during the correction, captured by the distortion factor $(1 - 2i\chi_{sb}\hat{a}^{\dagger}\hat{a} / \kappa_{b})$ and the added term in the Hamiltonian, can then significantly degrade the bit-flip performance of the cat qubit.

In practice, the storage self-Kerr and storage-buffer cross-Kerr originate from the self-Kerr of the buffer mode and the storage-buffer hybridization. In an ideal realization of the ATS buffer element, the self-Kerr of the buffer mode vanishes at the saddle points (i.e., first-order flux-insensitive points). However, as noted in previous work~\cite{Lescanne2020}, an ATS in practice typically has non-zero self-Kerr on the saddle points due to various imperfections, such as the finite number of junctions in the junction array and the asymmetry in the side junction energies. Here, we demonstrate that it is also important to consider inductances in series with the side junctions, as they can contribute significantly to the frequency and the self-Kerr of the buffer mode.

As shown in \cref{fig:serial_inductance}(a), the serial inductances introduce two extra flux nodes $\hat{\phi}_{P_{1}}$, $\hat{\phi}_{P_{2}}$ to the system, in addition to the flux node $\hat{\phi}$ of the buffer mode. We focus on the regime where these serial inductances are much smaller than the inductances due to the side junctions, i.e., $E_{L,P_{i}} = \hbar^{2} / (4e^{2}L_{P_{i}}) \gg E_{J,i}$ for $i=1,2$. In this regime, the extra nodes $\hat{\phi}_{P_{1}}$, $\hat{\phi}_{P_{2}}$ can be treated classically and set to be the values that minimize the inductive potential function~\cite{Quintana2017thesis, Kafri2017}. Then we obtain an effective Hamiltonian (see \cref{app:serial_inductance_calculations}), 
\begin{align}
    \hat{H} &= 4E_{C}\hat{N}^{2}  -NE_{J,\text{array}} \cos\Big{(}\frac{\hat{\phi}}{N}\Big{)} 
    \nonumber\\
    &  -E_{J,1}\cos(\hat{\phi} + \varphi_{x,1}) - E_{J,2}\cos(\hat{\phi} - \varphi_{x,2})  
    \nonumber\\
    & + \frac{E_{J,1}^{2}}{4E_{L, P_{1}}}\cos(2(\hat{\phi} + \varphi_{x,1}))   + \frac{E_{J,2}^{2}}{4E_{L, P_{2}}}\cos(2(\hat{\phi} - \varphi_{x,2})) , 
\end{align}
where the first two lines represent the Hamiltonian of an ATS as described in Ref.~\cite{Lescanne2020}, and $N$ is the number of junctions in the junction array. The last line represents the additional contributions from the serial inductances. Notably, these additional terms have a different flux periodicity from that of the usual ATS potential due to the extra factor of $2$ in the arguments of the cosine terms.  

\cref{fig:serial_inductance}(b) provides direct experimental evidence for the contributions of the serial inductances in the buffer mode of our device.  We show the frequency of the buffer mode (black curve) as a function of $\varphi_\Delta \equiv (\varphi_{x,1} - \varphi_{x,2})/2$ with  $\varphi_{\Sigma}\equiv (\varphi_{x,1} + \varphi_{x,2})/2 = \pi/2$. Notably, a fitted model without the serial inductances (blue curve) fails to capture the correction that is periodic in $\varphi_{\Delta}$ with period $\pi$ (this is in comparison to the $2\pi$-periodicity in an ideal ATS). On the other hand, the fitted model with the serial inductances (orange curve) captures this doubly-periodic behavior. From the latter model, which assumes $L_{P_{1}}=L_{P_{2}}=L_{P}$, we extract a serial inductance of $L_{P} = 27~\text{pH}$, corresponding to $E_{J,i} / E_{L,P_{i}} \lesssim 0.01$ in our device for $i=1,2$.

The effect of the serial inductances on the buffer frequency is useful for  characterizing the inductances but is itself not critical to the cat qubit performance. However the impact of serial inductances on the self-Kerr of the buffer mode is crucial. For example, assuming symmetry between the two side junctions, i.e., $E_{J,1} = E_{J,2} = E_{J}$ and $E_{L,P_{1}}=E_{L,P_{2}} = E_{L,P}$, we find that the buffer self-Kerr is perturbatively given by
\begin{align}
    K_{b} = -\frac{E_{C}}{N^{2}} + \frac{8E_{C}E_{J}^{2}}{E_{L,P}E_{L}}, \label{eq:buffer_self_Kerr}
\end{align}
on the saddle points (e.g., $(\varphi_{\Sigma},\varphi_{\Delta}) = (\pi/2, \pi/2)$ and $(\varphi_{\Sigma},\varphi_{\Delta}) = (\pi/2, 3\pi/2)$), and where $E_{L} \equiv E_{J,\text{array}} / N$ is the effective inductive energy of the junction array (see \cref{app:serial_inductance_calculations}). This expression shows that the serial inductances introduce positive self-Kerr on the buffer mode. 

Previous works have used a large number of junctions in the junction array (e.g., $N=5$~\cite{Lescanne2020,Berdou2023} or $N=13$~\cite{Reglade2024}) to minimize the negative self-Kerr contribution $-E_{C}/N^{2}$ from the junction array. On the other hand, our analysis suggests a different strategy. Given the serial inductance, we can optimize the circuit parameters of the buffer such that the negative self-Kerr from the junction array is canceled out by the positive self-Kerr from the serial inductances. In particular, this strategy allows us to use a shallow junction array with just $N=3$, reducing the required junction energy $E_{J,\text{array}}$ to achieve a given target $E_{L}$.    

In \cref{fig:serial_inductance}(c) we illustrate how nonlinearities can be minimized by carefully tuning the design value of $E_J$. We provide predictions of the storage self-Kerr, $K_{s}$, and the storage-buffer cross-Kerr, $\chi_{sb}$, as a function of $E_{J}$ using a model where the other underlying circuit parameters are tuned to match the buffer frequency data in \cref{fig:serial_inductance}(b). This model predicts that these buffer-induced storage-mode nonlinearities would vanish at an optimal average side junction energy of $E_{J}/h = 40$~GHz (blue dashed line) due to an ideal balancing of the positive and negative self-Kerr contributions on the buffer mode. For the device of this work, we extract an $E_{J}/h$ of $57\text{GHz}$ (black dashed line). Despite the offset from the optimal value, the predicted storage self-Kerr of $K_{s} / 2\pi = 1.1~\text{kHz}$ and the storage-buffer cross-Kerr of $\chi_{sb}/2\pi = 156~\text{kHz}$ are still tolerable given the strength of the 3WM interaction $g_{2} / 2\pi = 578\pm 4~\text{kHz}$ and the buffer decay rate $\kappa_{b}/2\pi = 10.7~\text{MHz}$. We independently measure the self-Kerr of the storage mode and find it to be $K_{s}/2\pi = 0.97\pm 0.02 ~\text{kHz}$ (green horizontal strip in \cref{fig:serial_inductance}(c); see also \cref{app:storage_kerr}), in good agreement with the model prediction. 

\section{Two-photon dissipation performance}
\label{sec:two_photon_dissipation_performance}

\begin{figure}[t!]
    \centering
    \includegraphics[width=\columnwidth]{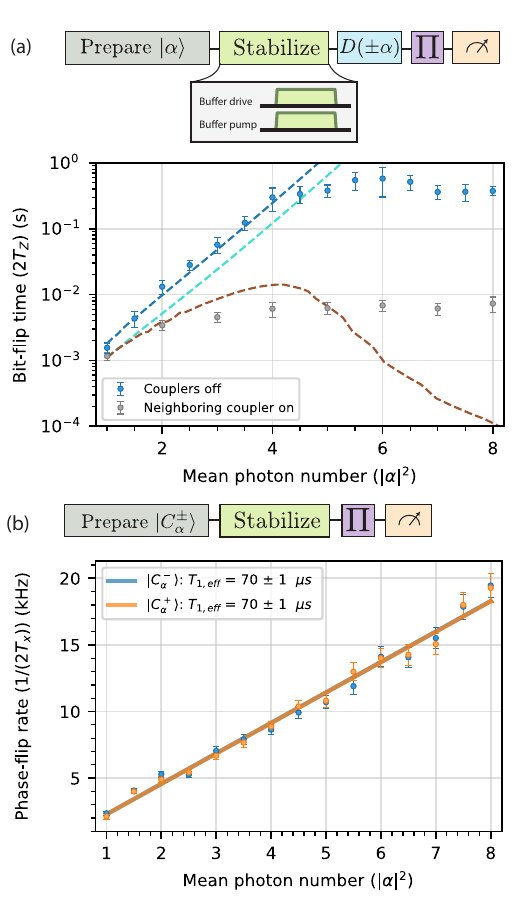}
    \caption{\textbf{Two-photon dissipation performance.} (a) Bit-flip times of our cat qubit as a function of the mean photon number.  The blue points are the bit-flip times of the cat qubit when all the couplers are in the off-position. In the gray curve we show the bit-flip times when the coupler in the neighboring unit cell is near its minimum frequency position ($\Phi/\Phi_0=0.4$). The dashed turquoise curve shows simulated bit-flip times based on the device parameters with couplers off. The dashed blue curve shows simulated bit-flip times where the white-noise dephasing parameter is phenomenologically tuned to match the experimental data. In the dashed brown curve we show a simulation with twice larger $E_J$ and $L_P$ of the buffer where significant degradation is observed due to buffer-induced nonlinearities.  (b) Phase-flip times of the cat qubit. Lines correspond to a linear fit to the phase-flip rate, $\gamma_{Z} = \kappa_{1,\text{eff}}|\alpha|^{2}$, where $T_{1,\text{eff}}=1/\kappa_{1,\text{eff}}$.}
    \label{fig:cat_lifetimes}
\end{figure}

Having presented the high coherence and low nonlinearity of the storage mode in \cref{sec:tunable_dispersive_coupling,sec:storage_mode_coherence,sec:storage_linearity}, we now proceed to characterize the performance of a cat qubit encoded in this mode. As shown in the diagram in \cref{fig:cat_lifetimes}(a), we characterize the bit-flip time of a cat qubit by initializing a coherent state $|+\alpha\rangle \simeq |0\rangle$ and then stabilizing the cat qubit manifold with a 3WM pump and a drive on the buffer mode. Finally, using the tunable coupler we perform parity measurements after $\pm\alpha$ displacements to read out the cat qubit in the $|0\rangle/|1\rangle$ basis ($Z$ basis). The bit-flip time $T_{\text{bit-flip}} = 2T_{Z}$ is characterized by fitting these $Z$-basis populations to an exponential curve, and then extracting the decay exponent $T_{Z}$ (see \cref{app:error_rate_convention}).   

The dark blue data points in \cref{fig:cat_lifetimes}(a) show the bit-flip time of our cat qubit increases exponentially in the mean photon number $|\alpha|^{2}$, and begins to saturate at a maximum value of $0.3~\text{s} - 0.6~\text{s}$ when $|\alpha|^{2} \gtrsim 4$. We fit the data for $|\alpha|^2\leq 4$ to the relation $T_{\text{bit-flip}}\propto e^{\gamma|\alpha|^{2}}/|\alpha|^{2}$ and find that the extracted exponent $\gamma = 2.24\pm 0.09$ is comparable with the optimal value of $\gamma=2$ expected under a white-noise dephasing model~\cite{mirrahimi2014} (larger values of $\gamma$ up to $\gamma=4$ are possible with only single-photon loss in the absence of dephasing~\cite{chamberland2020building,Dubovitskii2024}). A fit to a model without the factor of $|\alpha|^2$, $T_{\text{bit-flip}}\propto e^{\gamma' |\alpha|^{2}}$, which provides more direct intuition on the rate of scaling with photon number, yields $\gamma' =1.75 \pm 0.06$. This corresponds to a factor of $5.8$ increase in the bit-flip time for every added photon. Owing to the large scaling exponent and long storage-mode coherence times, we achieve a bit-flip time over $1~\text{ms}$ with only $|\alpha|^{2}=1$, which increases to above $0.1~\text{s}$ with just $|\alpha|^{2}=4$.  

The dashed turquoise curve in \cref{fig:cat_lifetimes}(a) shows the simulated bit-flip time based on the measured storage-mode coherence times in \cref{sec:storage_mode_coherence} and the predicted (and measured) storage-mode nonlinearities in \cref{sec:storage_linearity}. The simulation agrees approximately with the experimental data (up to a multiplicative offset) in the $|\alpha|^{2} \le 4$ regime, where the exponential bit-flip suppression is observed. The simulation with tuned white-noise dephasing parameter (dashed blue curve; see \cref{app:effective_model_analysis}) more closely captures the experimental data without the multiplicative offset. Finally, to show the importance of carefully selecting buffer parameters, we perform an additional simulation with the buffer mode having twice as large side junction energy $E_J$ and serial inductance $L_{P}$ (dashed brown curve). The simulation predicts poor bit-flip performance which is primarily due to the intolerably large storage self-Kerr and storage-buffer cross-Kerr (approximately $8$ times larger than the device parameters). This demonstrates the importance of understanding the contributions of the buffer serial inductances to the nonlinearities of the storage-buffer system.  

To confirm the effectiveness of the extinction of the coupling provided by the tunable coupler, we perform an experiment where we apply dissipative stabilization while we purposefully turn on the coupler between the storage mode and the transmon ancilla in a neighboring unit cell of our device. When the external flux of this one coupler is set to $0.4\Phi_{0}$ (i.e., close to the on-position), the bit-flip time of the cat qubit under dissipative stabilization saturates at a value below $10~\text{ms}$, as indicated by the gray data points in \cref{fig:cat_lifetimes}(a). This saturation in bit flip time is due to heating of the coupler and the ancilla in the neighboring unit cell. With the coupler in the on-position, a heating event in either the coupler or the ancilla gives rise to a dispersive shift in the storage mode frequency larger than that of the ($1~\text{MHz}$-level) confinement rate of the dissipative stabilization over the range of mean photon number studied here. This leads to a loss of confinement of the storage mode, and a high probability of a cat-qubit bit flip with each heating event~\cite{Lescanne2020}. With all of the couplers at their off-position, the bit-flip time far exceeds the ancilla and coupler heating rate limit of $10~\text{ms}$, but eventually saturates at a maximum value of $0.3~\text{s} - 0.6~\text{s}$. We have not identified a definitive mechanism explaining this saturation behavior, and it remains an area of continued study (see \cref{app:dissipation_with_excitations} for more details).  

While the bit-flip rates of a cat qubit are (ideally) suppressed exponentially with $|\alpha|^{2}$, the phase-flip rates of a cat qubit are expected to increase linearly, i.e., $\gamma_{Z} = \kappa_{1,\text{eff}}|\alpha|^{2}$ where $\kappa_{1,\text{eff}}$ is the effective single-photon loss rate.  As shown in the diagram in \cref{fig:cat_lifetimes}(b), we characterize  phase-flip rates of our cat qubit by initializing the storage mode into an even or odd cat state by displacing the storage mode and performing a parity measurement.  Then we stabilize the cat qubit with the two-photon dissipation and finally perform a parity measurement to detect parity flips.  The phase-flip rate is then extracted through an exponential fit of the parity decay curve. The results in \cref{fig:cat_lifetimes}(b) show that the phase-flip rate increases linearly in the mean-photon number $|\alpha|^{2}$ as expected. From linear fits to this curve, we find that the effective phase-flip lifetime is $T_{1,\text{eff}} = 1/\kappa_{1,\text{eff}} = 70\pm 1~\mathrm{\mu s}$. This value is consistent with an independent measurement of the storage mode energy decay time of $T_{1} = 74\pm 1~\mathrm{\mu s}$ (see \cref{app:storage_t1_t2_msmts}). 

The results in this section demonstrate that our cat qubit is capable of achieving nearly ideal exponential bit-flip suppression in the regime of $|\alpha|^{2} \lesssim 4$, where the phase-flip rates remain relatively low.  In particular, we estimate that for a typical error-correction cycle length of $\sim 2~\mathrm{\mu s}$, the phase-flip probabilities of our cat qubit are approximately $3\%$ at $|\alpha|^2=1$ and $11\%$ at $|\alpha|^2=4$. Since the error threshold of a repetition code is approximately $11\%$~\cite{Dennis2002} (assuming a phenomenological noise model), the phase-flip error rates achieved are sufficiently low to be corrected by a repetition code in the regime where $|\alpha|^{2} \lesssim 4$. At the same time, we observe a large noise bias ranging from $10$ to over $1000$ across this same range of average photon numbers. Thus the observed phase-flip and bit-flip times meet the requirements for more complex operations with cat qubits, such as multi-qubit gates and error-correction experiments with an outer repetition code. 

\section{Pulsed Stabilization}
\label{sec:pulsed_stabilization}

\begin{figure}[t!]
    \centering
    \includegraphics[width=\columnwidth]{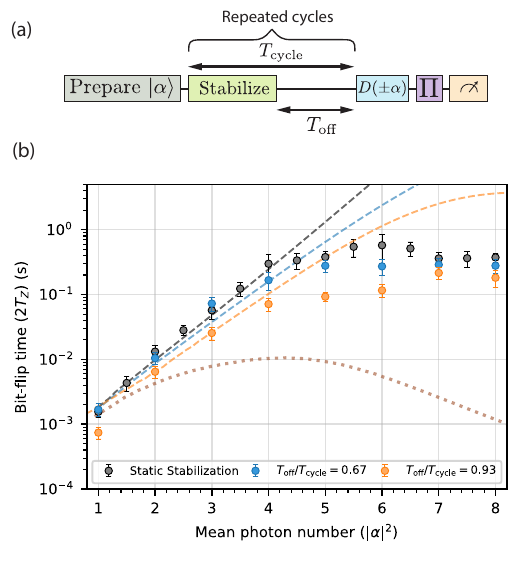}
    \caption{\textbf{Pulsed stabilization.}  (a) A high-level depiction of the measurement sequence for pulsed cat-qubit stabilization. Unlike previous experiments where we have the two-photon dissipation continually turned on, here we repeatedly apply pulsed stabilization cycles. In each cycle of length $T_{\text{cycle}}$, the stabilization is turned off for a duration $T_{\text{off}}$.  (b) Bit-flip times of our cat qubit under pulsed stabilization as a function of the mean photon number $|\alpha|^{2}$ for various duty cycle ratios of $T_{\text{on}} / T_{\text{cycle}}$ for $T_\text{cycle}=3~\mathrm{\mu s}$. The markers correspond to the measured data of a baseline static stabilization (black), pulsed stabilization with $T_{\mathrm{on}}=1~\mathrm{\mu s}$ (blue), and pulsed stabilization with $T_{\mathrm{on}}=200$~ns (orange).  The dashed lines correspond to simulations for each of the three cases.   The dotted brown line corresponds to a simulation with $T_{\mathrm{on}}=1~\mathrm{\mu s}$ where the storage self-Kerr is increased to $K_{s}/2\pi = 5~\text{kHz}$, corresponding to a value five times that of the self-Kerr predicted and measured for the device.  
    }
    \label{fig:pulsed_stabilization}
\end{figure}

In error-correction architectures involving cat qubits, it is often required to turn off two-photon dissipation on a cat qubit for certain syndrome-extraction operations. For example, implementation of a CNOT gate with a cat qubit as the target typically relies on rotating the target cat qubit conditioned on the state of a control qubit~\cite{Guillaud2019Repetition,Puri2020_bias,chamberland2020building,Gautier2022,putterman2024}.  For this conditional rotation to work well for a dissipative cat qubit one either needs to engineer a complex two-photon dissipation scheme with its stabilized manifold rotating conditioned on the state of a control qubit~\cite{Guillaud2019Repetition, chamberland2020building}, or turn off the two-photon dissipation such that the target cat qubit can freely rotate under a separate mechanism~\cite{Gautier2022,putterman2024}. In the latter case, it is essential to ensure that the bit-flip performance of the cat qubit is not significantly degraded when the two-photon dissipation is turned off for some fraction of the time during a cycle. This is a challenge because storage-mode errors due to mechanisms such as undesired nonlinearities and dephasing are allowed to freely accumulate when two-photon dissipation is turned off. 

In this section, we study the bit-flip performance of our cat qubit under pulsed stabilization. That is, as shown in \cref{fig:pulsed_stabilization}(a), we initialize a coherent state and then repeatedly apply a cycle of pulsed stabilization with cycle time $T_{\text{cycle}}$ in which the two-photon dissipation is pulsed on and off for a duration of $T_{\text{on}}$ and $T_{\text{off}} = T_{\text{cycle}} - T_{\text{on}}$, respectively. Then we characterize the bit-flip rate of our cat qubit by measuring the displaced parities similarly as in \cref{sec:two_photon_dissipation_performance}.  When calibrating the pulsed stabilization we ensure that the phase of the storage-mode is carefully tracked so that two-photon dissipation is aligned with the storage mode coherent states when it turns back on. 

In \cref{fig:pulsed_stabilization}(b), we show the bit-flip time of our cat qubit as a function of the mean photon number $|\alpha|^{2}$ for various duty cycle ratios, $T_{\text{on}} / T_{\text{cycle}}$, with a fixed cycle time of $T_{\text{cycle}} = 3$~$\mu$s. We include the baseline case of static stabilization where the two-photon dissipation is continually turned on (black). Relative to the baseline we observe no degradation in the bit-flip time even when the dissipation is turned on for only $1~\mathrm{\mu s}$, or one third of the time during the pulsed stabilization cycle (blue curve). Note that, unsurprisingly, the bit-flip times do not exceed the saturated value observed in the static stabilization case. Significant degradation of the cat-qubit bit-flip times under pulsed stabilization is observed only in an extreme limit of $T_{\text{on}}/T_{\text{cycle}}=7\%$, when the pulsed stabilization is turned on for only $T_{\text{on}} = 200$~ns of the $3~\mathrm{\mu s}$ cycle length.   

The dashed curves correspond to simulations of the pulsed stabilization procedure for different duty cycles of stabilization, $T_{\text{on}}/T_{\text{cycle}}$. These simulations use the same refined value for the white-noise dephasing parameter as in the dashed blue curve in \cref{fig:cat_lifetimes}(a). For all but the dotted brown curve, we also assume in these simulations the device estimated self-Kerr of $K_{s}/2\pi = 1.1$~kHz and storage-buffer cross-Kerr of $\chi_{sb}/2\pi = 156$~kHz.  We find that for a duty cycle of $T_{\text{on}}/T_{\text{cycle}}=1/3$ (blue), the degradation in the bit-flip times is relatively marginal, consistent with our measurements. When the pulsed stabilization is reduced to a $7\%$ duty cycle (orange), the simulations predict a more significant reduction in the bit-flip time relative to the static stabilization case (black), especially at large values of mean photon number $|\alpha|^{2}\gtrsim 6$. Although this is consistent with our measured observations, the simulation model does not capture the previously discussed, and currently unexplained, measured saturation in bit flip time at $0.3-0.6$~s. The dotted brown curve corresponds to an additional simulation of the case $T_{\text{on}}/T_{\text{cycle}}=1/3$, where we increase the storage self-Kerr nonlinearity to $K_{s}/2\pi = 5$~kHz (the cross-Kerr is left at $\chi_{sb}/2\pi = 156$~kHz). This self-Kerr is five times larger than what we predict and measure in our device (see \cref{sec:storage_linearity}), but representative if the buffer-mode nonlinearities are not carefully minimized. In this case, we compute that the bit-flip times are significantly degraded and remain below $\sim 10~\text{ms}$ due to the larger distortion caused by the increase in storage self-Kerr. This reinforces the importance of the accurate buffer modeling in \cref{sec:storage_linearity}, and the inclusion of serial inductances in the ATS, which ensure that the buffer-induced storage-mode nonlinearities stay in check. 

\section{Conclusion}

The work presented here sets the stage for performing complex multi-qubit gates and error-correction experiments involving dissipatively-stabilized cat qubits. The low buffer-mode nonlinearities and high storage-mode coherence allow us to achieve long cat-qubit bit-flip times with small mean photon numbers. For typical error correction cycle times, the phase-flip error probability of our cat qubit can stay below the error threshold of an outer repetition code up to the mean photon number of $|\alpha|^{2}\simeq 4$, where bit-flip times exceed $0.1~\text{s}$. We can also retain the benefits of exponentially suppressed bit-flip error rates even when the stabilization is turned off for a significant fraction of a pulsed-stabilization cycle.  Additionally, the cat qubits of this work are implemented using planar superconducting quantum circuits, and are amenable to scalable microfabrication processes. This scalability naturally allows for concatenating many such cat qubits into an outer error-correcting code, a compelling architecture for reducing the resource overhead associated with quantum error correction~\cite{Guillaud2019Repetition,Guillaud2021_error,Darmawan2021,chamberland2020building,Regent2023highperformance,Gouzien2023,Ruiz2024}.  

Going forward, it will be important to drive the noise bias of cat qubits many orders of magnitude higher, from the $10^3$ level of this work to levels approaching $10^7$, in order to fully realize the potential for hardware-efficient quantum error correction at algorithmically-relevant error rates~\cite{putterman2024}. Crucially the cat-qubit phase-flip rates still need to remain sufficiently low even at larger cat-qubit mean photon numbers needed for realizing such a large noise bias. Thus it will also be important to push storage-mode $T_{1}$ lifetimes to $1$~ms and beyond to accommodate larger values of cat-qubit mean photon number. Critical to this pursuit, the strategies presented in this work can be used to accommodate higher storage coherence times and achieve stronger two-photon dissipation rates without introducing parasitic effects. Future work to uncover and mitigate the mechanism(s) responsible for the observed saturation of bit-flip times at the $1$-second level in our current devices will also be important. More generally, pushing the limits of cat qubit bit-flip times and noise bias will open up new opportunities to study error mechanisms in superconducting quantum circuits that are rare events and occur on long timescales~\cite{McEwen2022, Acharya2024}.

\begin{acknowledgments}

We thank Alex Retzker and Yufeng Ye for helpful comments on the manuscript, and the staff from across the AWS Center for Quantum Computing that enabled this project. We also thank Fiona Harrison, Harry Atwater, David Tirrell, and Tom Rosenbaum at Caltech, and Simone Severini, Bill Vass, James Hamilton, Nafea Bshara, and Peter DeSantis at AWS, for their involvement and support of the research activities at the AWS Center for Quantum Computing. 

\end{acknowledgments}

\appendix

\section{Device fabrication}
The cat qubit system described in the main text is implemented as one unit cell of a larger integrated device. The device consists of two chip dies which are fabricated separately on high-resistivity silicon and then flip-chip bonded together~\cite{Foxen2018, Das2018}. The first (“qubit”) chip die contains aluminum metallization layers, including Al/AlOx/Al Josephson junctions which form the nonlinear circuit elements in the transmon ancilla and couplers, as well as the ATS buffer. The Josephson junctions leads, deposited with a double-angle evaporation process, are respectively shorted to the aluminum ground plane and island capacitors with an added “bandage” contact~\cite{Dunsworth2017, Keller2017}. The second chip die is metallized with thin-film alpha phase tantalum (thickness $\sim 100~\text{nm}$), which has lower microwave losses than aluminum and is capable of high-coherence storage modes in coplanar waveguide resonators~\cite{Place2021,Crowley2023}. The patterned circuit consists of storage resonators and other linear elements such as the metamaterial filter and readout resonators. Decoupling fabrication of the constituent dies with separate process flows allows us to take advantage of tantalum for achieving high-coherence storage modes without the constraints imposed by integrating Al/AlOx/Al Josephson junctions on the same chip die.

\section{Wiring diagram and experimental setup}

\begin{figure*}[t!]
    \centering
    \includegraphics[width=1.85\columnwidth]{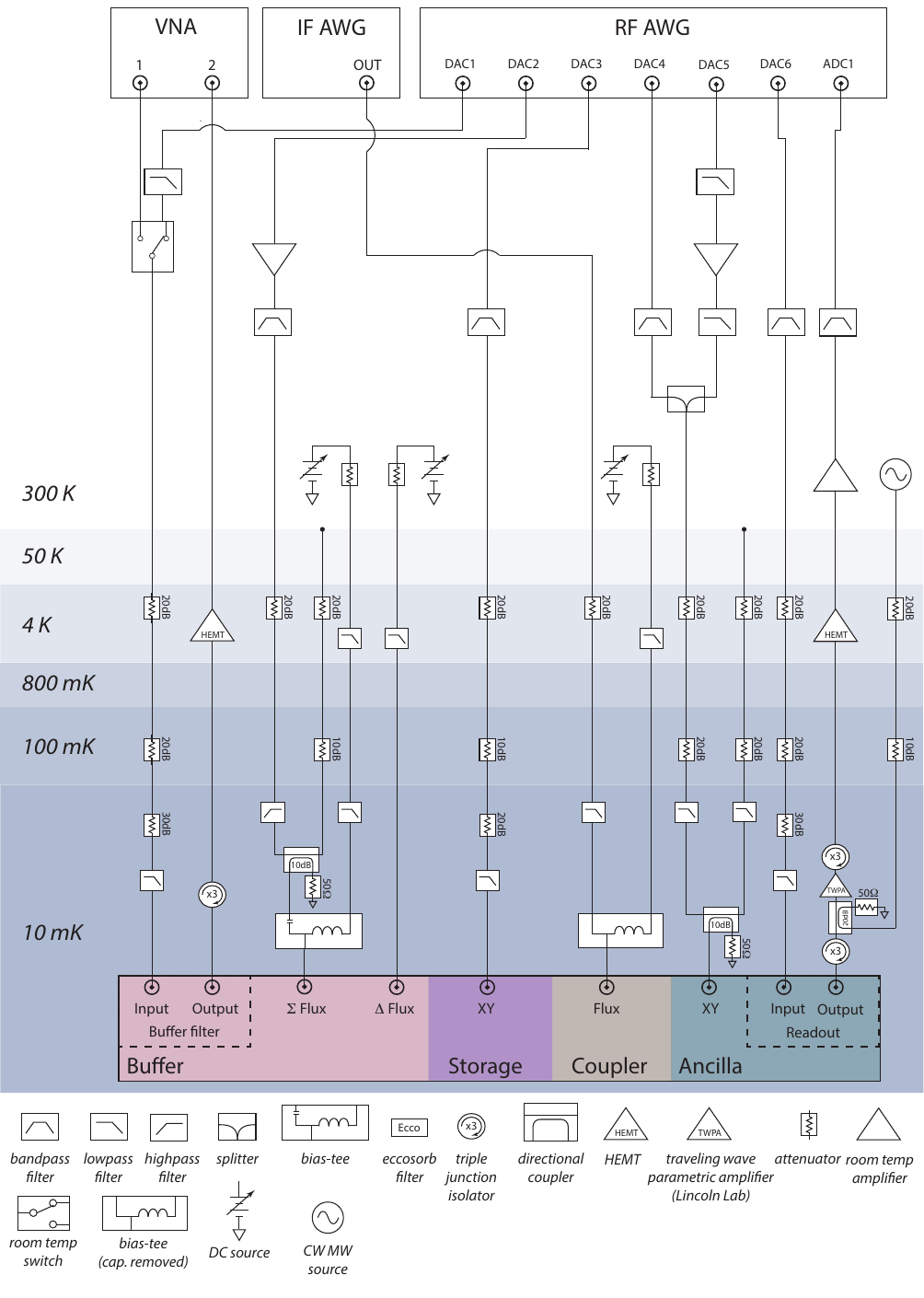}
    \caption{\textbf{Wiring diagram.}  Some details not included are eccosorb filters on almost all of the lines and room temperature switches to toggle between characterization of different components.}
    \label{app_fig:wiring_diagram}
\end{figure*}

A diagram including the fridge wiring and room temperature setup for one unit cell is shown in \cref{app_fig:wiring_diagram}.  

\section{Table of device properties}
\label{app:device_parameters}

Tables containing the device properties are given in \cref{table:comprehensive_device_properties,table:additional_device_properties}. 

\begin{table}[h!]
\begin{center}
\begin{tabular}{
|c c c c|} 
\hline
Property & Descriptor (Reference) & Unit & Value \\ [0.5ex] 
\hline

\multicolumn{4}{|l|}{Buffer properties} \\
$\omega_b/2\pi$ & Buffer frequency & GHz & 3.01 \\ 
$\kappa_b/2\pi$ & Buffer loss rate & MHz & 10.7  \\ 
$(\omega_b - \omega_d)/2\pi$ & Buffer freq. - drive freq. & MHz & 29.3  \\ 
\hline

\multicolumn{4}{|l|}{Coupler maximum frequency (idle) position properties}  \\
$\Phi_{x,c}/\Phi_0$ & Coupler flux value & & 0 \\
$\omega_{c}/2\pi$ & Coupler frequency & GHz & 8.40  \\
$\omega_a/2\pi$ & Ancilla frequency & GHz & 5.21  \\
$\omega_s/2\pi$ & Storage frequency & GHz & 5.35  \\
$\chi_{sa}/2\pi$ & Storage-ancilla cross-Kerr & kHz & -1.4  \\
$\omega_r/2\pi$ & Ancilla readout frequency & GHz & 7.23  \\ 
$K_s/2\pi$  & Storage Kerr & kHz & 0.97  \\ 
$(\omega_{s} - \bar{\omega}_{s})/2\pi$  & Storage Stark shift under 3WM & MHz & 0.39  \\ 
\hline

\multicolumn{4}{|l|}{Parity measurement position properties}  \\
$\Phi/\Phi_0$ & Coupler flux value & & 0.38 \\
$\chi_{sa}/2\pi$ & Storage-ancilla cross-Kerr & MHz & -0.9  \\
\hline

\multicolumn{4}{|l|}{Number splitting position properties}  \\
$\Phi/\Phi_0$ & Coupler flux value & & 0.43 \\
$\chi_{sa}/2\pi$ & Storage-ancilla cross-Kerr & MHz & -2.4  \\
\hline

\multicolumn{4}{|l|}{Coupler minimum frequency position properties}  \\
$\Phi/\Phi_0$ & Coupler flux value & & 0.5 \\
$\omega_{c}/2\pi$ & Coupler frequency & GHz & 5.77  \\
$\chi_{sa}/2\pi$ & Storage-ancilla cross-Kerr & MHz & -5.7  \\

\hline
\end{tabular}
\end{center}
\caption{Table of device properties for the unit cell used to characterize the storage mode.}  
\label{table:comprehensive_device_properties}
\end{table}

\begin{table}[h!]
\begin{center}
\begin{tabular}{
|c c c c|} 
\hline
Property & Descriptor (Reference) & Unit & Value \\ [0.5ex] 
\hline

\multicolumn{4}{|l|}{Coupler maximum frequency (idle) position properties}  \\
$\Phi_{x,c}/\Phi_0$ & Coupler flux value & & 0 \\
$\omega_{c}/2\pi$ & Coupler frequency & GHz & 8.9  \\
$\omega_a/2\pi$ & Ancilla frequency & GHz & 5.2  \\
$\chi_{sa}/2\pi$ & Storage-ancilla cross-Kerr & kHz & -2.0  \\
\hline

\multicolumn{4}{|l|}{Coupler minimum frequency position properties}  \\
$\Phi/\Phi_0$ & Coupler flux value & & 0.5 \\
$\omega_{c}/2\pi$ & Coupler frequency & GHz & 5.8  \\
$\chi_{sa}/2\pi$ & Storage-ancilla cross-Kerr & MHz & -4.3  \\

\hline
\end{tabular}
\end{center}
\caption{Abridged table of device properties for the additional unit cell containing the coupler used in \cref{fig:cat_lifetimes}(a). }
\label{table:additional_device_properties}
\end{table}

\section{Calibration details}
The buffer is flux biased to a saddle point and the couplers neighboring the storage mode are flux biased to their maximum frequency positions.  To calibrate the dispersive interaction between the ancilla and storage we first select a flux pulse amplitude for the coupler.  We initially start with a flux pulse amplitude which brings the coupler to its minimum frequency position.  We calibrate the interaction frequency of the ancilla which is shifted from the idling ancilla frequency due to the increased hybridization with the coupler.  With these calibrations completed, we have the capability to excite the ancilla in a storage-photon-number-selective manner. Using the photon-number-selective ancilla pulses we perform the initial calibration of the storage interaction frequency and flux pulse lengths using the storage conditional phase experiment described in \cref{app:storage_conditional_phase}. For parity measurements the ancilla phase must also be carefully tracked during the flux pulse so we calibrate an ancilla phase correction used for this purpose.

Using the coupler's minimum frequency position is impractical in our system as the strong couplings make the dispersive nature of the storage-ancilla interaction break down (see \cref{app:tunable_dispersive_coupling_details}). We thus calibrate a few additional operating points of the coupler besides its minimum frequency position including the ``parity measurement'' and ``number splitting'' positions whose parameters are provided in \cref{table:comprehensive_device_properties}.  The ``parity measurement'' position is used for Wigner tomography and the final storage-mode measurements in bit-flip and phase-flip measurements in \cref{fig:cat_lifetimes}.  The ``number splitting'' position is used in the number splitting measurements of \cref{fig:tunable_coupler} and $T_2$ measurements in \cref{fig:storage_coherence}. Calibration of these additional operating points follows a similar procedure used for calibrating the minimum frequency position. For the number splitting measurements, which take longer than parity measurements and are more sensitive to flux pulse nonidealities (for example long timescale distortions), we fine tune the ancilla interaction frequency with a spectroscopy measurement.  

For both the parity measurements and number-selective measurements, we use flux pulses with net-zero~\cite{DiCarlo2021} waveforms with Gaussian shoulders.  The number-selective ancilla pulses themselves have a Gaussian shape with a pulse length of $1.4~\mathrm{\mu s}$ and are applied during the second half of the net-zero flux pulse. Once parity measurements are calibrated we can calibrate the storage mode displacement as explained in \cref{app:displacement_dissipation_amplitude_calibration}. 

For the initial two-photon dissipation calibration we measure the buffer 3WM condition using a vector network analyzer.  The two-photon dissipation is fine tuned to match the Stark-shifted storage frequency under the 3WM buffer pump by performing a storage Ramsey experiment with a pure two-photon dissipation applied. This calibration is performed frequently to ensure that the pump and drive frequencies of the two-photon dissipation stay up to date with the Stark-shifted storage mode frequency (i.e., $(\omega_{p} + \omega_{d})/2 \simeq \bar{\omega}_{s}$). 

The buffer drive frequency ($\omega_d$) should ideally match the Stark-shifted buffer frequency ($\bar{\omega}_b$). In practice, the two-photon dissipation performs comparably with an appreciable buffer detuning ($\bar{\omega}_b-\omega_d$) on the order of $1~\text{MHz}$, which is to be compared against the buffer decay rate $\kappa_{b}$. As a result we only coarsely optimize the buffer drive frequency to $1~\text{MHz}$-level precision.  The calibration of the linear relationship between the average photon number $|\alpha|^{2}$ of the cat qubit and the buffer drive amplitude is described in \cref{app:displacement_dissipation_amplitude_calibration}.

\section{Basis convention}
\label{app:basis_convention}

We define the complementary basis states $|\pm\rangle$ of the cat qubit as the even and odd cat states, i.e., 
\begin{align}
    |\pm\rangle \equiv |C^{\pm}_{\alpha}\rangle = \frac{|+\alpha\rangle \pm |-\alpha\rangle}{\sqrt{2(1 \pm e^{-2|\alpha|^{2}})}}  . 
\end{align}
Then, through the Hadamard transformation, the computational basis states approach the coherent states $|\pm\alpha\rangle$ in the large $|\alpha|$ limit where $e^{-2|\alpha|^{2}}$ becomes negligible: 
\begin{align}
    |0\rangle  &\equiv \frac{1}{\sqrt{2}}(|+\rangle + |-\rangle) \xrightarrow{ e^{-2|\alpha|^{2}} \ll 1 }  |+\alpha\rangle, 
    \nonumber\\
    |1\rangle  &\equiv \frac{1}{\sqrt{2}}(|+\rangle - |-\rangle) \xrightarrow{ e^{-2|\alpha|^{2}} \ll 1 }  |-\alpha\rangle. 
\end{align}
Note that in the opposite limit where $|\alpha|$ approaches zero, we have 
\begin{align}
    |+\rangle & \xrightarrow{|\alpha| \rightarrow 0} |\hat{n}=0\rangle, 
    \nonumber\\
    |-\rangle & \xrightarrow{|\alpha| \rightarrow 0} |\hat{n}=1\rangle, 
\end{align}
where $|\hat{n}=0\rangle$ and $|\hat{n}=1\rangle$ are the vacuum and single-photon states, respectively (i.e., eigenstates of the number operator $\hat{n} = \hat{a}^{\dagger}\hat{a}$). This implies that in the $|\alpha|\rightarrow 0$ limit, the computational basis states of a cat qubit are given by 
\begin{align}
    |0\rangle &\equiv \frac{1}{\sqrt{2}}(|+\rangle + |-\rangle) = \frac{1}{\sqrt{2}}(|\hat{n}=0\rangle + |\hat{n}=1\rangle), 
    \nonumber\\
    |1\rangle &\equiv \frac{1}{\sqrt{2}}(|+\rangle - |-\rangle) = \frac{1}{\sqrt{2}}(|\hat{n}=0\rangle - |\hat{n}=1\rangle), 
\end{align}
i.e., equal superposition of the vacuum and single-photon states with the $\pm 1$ phases.  

In this basis convention, a single photon loss operator $\hat{a}$ maps the $|+\rangle$ state into the $|-\rangle$ state and vice versa, leading to phase-flip ($Z$) errors. A Pauli-$X$ measurement (i.e., $|+\rangle/|-\rangle$ basis measurement) of a cat qubit can be performed by reading out the photon-number parity $\exp[i\pi\hat{n}]$ since the $+1$ (or $-1$) eigenstate of $\hat{X}$ has even (or odd) number of photons. Since the computational basis states are approximately given by two coherent states $|\pm\alpha\rangle$ in the large $|\alpha|$ limit, a Pauli-$Z$ measurement (i.e., $|0\rangle/|1\rangle$ basis measurement) can be performed by measuring a displaced photon-number parity of the storage mode \cite{touzard2018,Lescanne2020}. 

\section{Error rate convention}
\label{app:error_rate_convention}

To define error rates and decay times of a cat qubit, we consider a simple Lindblad master equation of the form
\begin{align}
    \frac{d\hat{\rho}(t)}{dt} &= \gamma_{X}\mathcal{D}[\hat{X}]\hat{\rho}(t) + \gamma_{Y}\mathcal{D}[\hat{Y}]\hat{\rho}(t) + \gamma_{Z}\mathcal{D}[\hat{Z}]\hat{\rho}(t) , 
\end{align}
where $\hat{\rho}(t)$ is a qubit density operator, and $\hat{X},\hat{Y},\hat{Z}$ are the qubit Pauli operators. Under this master equation, an initial qubit state $\hat{\rho}(0)$ is depolarized by a Pauli error channel 
\begin{align}
    \hat{\rho}(t) &= (1 - p_{X}(t) - p_{Y}(t) + p_{Z}(t))\hat{\rho}(0) 
    \nonumber\\
    &\quad + p_{X}(t)\hat{X}\hat{\rho}(0)\hat{X} + p_{Y}(t)\hat{Y}\hat{\rho}(0)\hat{Y} + p_{Z}(t)\hat{Z}\hat{\rho}(0)\hat{Z} ,  
\end{align}
where the Pauli error probabilities are given by 
\begin{align}
    p_{X}(t) &= \frac{1}{4}(1 - e^{-2(\gamma_{X}+\gamma_{Y})t} + e^{-2(\gamma_{Y}+\gamma_{Z})t} - e^{-2(\gamma_{X}+\gamma_{Z})t}), 
    \nonumber\\
    p_{Y}(t) &= \frac{1}{4}(1 - e^{-2(\gamma_{X}+\gamma_{Y})t} - e^{-2(\gamma_{Y}+\gamma_{Z})t} + e^{-2(\gamma_{X}+\gamma_{Z})t}), 
    \nonumber\\
    p_{Z}(t) &= \frac{1}{4}(1 + e^{-2(\gamma_{X}+\gamma_{Y})t} - e^{-2(\gamma_{Y}+\gamma_{Z})t} - e^{-2(\gamma_{X}+\gamma_{Z})t}).  
\end{align}
Then it follows that the expectation values of the Pauli $\hat{Z}$ and $\hat{X}$ operators decay exponentially in time, i.e.,
\begin{align}
    \langle \hat{Z}\rangle(t) &= e^{-2(\gamma_{X} + \gamma_{Y})t} \langle \hat{Z}\rangle(0), 
    \nonumber\\
    \langle \hat{X}\rangle(t) &= e^{-2(\gamma_{Z} + \gamma_{Y})t} \langle \hat{Z}\rangle(0) . 
\end{align}
These observables $\langle \hat{Z}\rangle(t)$ and $\langle \hat{X}\rangle(t)$ are directly measurable in experiments. Specifically, we measure $\langle \hat{Z}\rangle(t)$ based on the population difference in the two coherent states $|\pm\alpha\rangle$ which are measured via displaced parity measurements. Moreover, $\langle \hat{X}\rangle(t)$ is measured through a parity measurement. Then, we fit the measured $\langle \hat{Z}\rangle(t)$ and $\langle \hat{X}\rangle(t)$ curves to an exponential decay curve and define $T_{Z}$ and $T_{X}$ as the decay constants. For $\langle \hat{Z}\rangle(t)$, we do not include a constant offset in the fit (i.e., $\langle \hat{Z}\rangle(t) = A \exp[ - t / T_{Z}]$) whereas for $\langle \hat{X}\rangle(t)$ a constant offset is included (i.e., $\langle \hat{X}\rangle(t) = A \exp[- t / T_{X}] + B$). A constant offset is needed in the latter case because in the small $|\alpha|^{2}$ regime (e.g., $|\alpha|^{2}=1$), the steady-state parity $\langle \hat{X}\rangle(t \rightarrow \infty)$ is non-zero due to asymmetric phase-flip rates between the $|+\rangle \rightarrow |-\rangle$ and $|-\rangle \rightarrow |+\rangle$ transitions (e.g., in the extreme limit of $|\alpha|^{2}=0$, $|+\rangle\rightarrow |-\rangle$ is unlikely to happen because $|+\rangle$ is given by the vacuum state but the single-photon state $|-\rangle$ can easily decay to the vacuum state $|+\rangle$).  

For a stabilized cat qubit with $|\alpha|^{2} \gg 1$, we anticipate that the noise model is well described by the above Lindblad master equation with $\gamma_{Z} \gg \gamma_{X} \gg \gamma_{Y}$ due to the biased noise structure. Thus, we approximately have 
\begin{align}
    T_{Z} = \frac{1}{2(\gamma_{X} + \gamma_{Y})} \simeq \frac{1}{2\gamma_{X}}, \quad T_{X} = \frac{1}{2(\gamma_{Z}+\gamma_{Y})} \simeq \frac{1}{2\gamma_{Z} } , \label{app_eq:bit_and_phase_flip_time_and_TZ_TX}
\end{align}
and the Pauli error probabilities are given by
\begin{align}
    p_{X}(t) \simeq \gamma_{X}t, \quad p_{Z}(t) \simeq \gamma_{Z}t , \label{app_eq:linearized_Pauli_error_probabilities} 
\end{align}
in the short time limit (i.e., $\gamma_{Y}t \ll \gamma_{X}t, \gamma_{Z}t \ll 1$). Based on the relation in \cref{app_eq:linearized_Pauli_error_probabilities}, we refer to $\gamma_{X}$ and $\gamma_{Z}$ as the bit-flip and phase-flip rates, i.e., added bit-flip and phase-flip probability per unit time. Correspondingly, we define the bit-flip and phase-flip times to be 
\begin{align}
    T_{\text{bit-flip}} &= \frac{1}{\gamma_{X}} = 2T_{Z},  
    \nonumber\\
    T_{\text{phase-flip}} &= \frac{1}{\gamma_{Z}} = 2T_{X}, 
\end{align}
where the last equality in each line is due to \cref{app_eq:bit_and_phase_flip_time_and_TZ_TX}. Thus, in our error rate convention, the bit-flip and phase-flip times are given by twice the decay times of the measured $\langle \hat{Z}\rangle(t)$ and $\langle \hat{X}\rangle(t)$ curves, respectively. 

\section{Device Hamiltonian}
In the rotating frame, the desired effective Hamiltonian of our system is given by 
\begin{align}
    \hat{H} &= \Big{(}g_{2}(t)(\hat{a}^{2} - \alpha^{2})\hat{b}^{\dagger} + \text{H.c.}\Big{)} + \chi_{sa}(\Phi_{x,c}(t))\hat{a}^{\dagger}\hat{a} |e_{a}\rangle\langle e_{a}|, 
\end{align}
where $\hat{a}$ and $\hat{b}$ are the annihilation operators of the storage and the buffer and $|e_{a}\rangle$ is the first excited state of the ancilla transmon. $|\alpha|^{2}$ is approximately the mean photon number of a cat qubit. $g_{2}(t)$ is the strength of the desired three-wave mixing (3WM) mixing process which is a crucial ingredient of the two-photon dissipation process. $|g_{2}(t)|$ is controlled by the amplitude of the sigma flux pump on the buffer mode. The 3WM process is continuously turned on (i.e., $|g_{2}(t)| > 0$) in ``static'' stabilization of a cat qubit (\cref{sec:two_photon_dissipation_performance}) and is periodically turned on and off in a pulsed stabilization scenario (\cref{sec:pulsed_stabilization}). In both cases, the coupler flux is nominally set to $\Phi_{x,c}(t) = 0$ (i.e., parking the coupler at its maximum frequency) to minimize the storage-ancilla dispersive coupling $\chi_{sa}$ while the cat qubit is being stabilized. Finally, at the end of an experimental sequence, the 3WM mixing process is turned off (i.e., $g_{2}(t) = 0$) and a flux pulse is applied to the coupler (i.e., $|\Phi_{x,c}(t)| > 0$) to turn on the storage-ancilla dispersive coupling $\chi_{sa}$. This dispersive coupling enables the characterization of the storage mode through the ancilla transmon (\cref{sec:tunable_dispersive_coupling}).    

Notable imperfections that are not included in the desired effective Hamiltonian above are the storage-buffer cross-Kerr $\chi_{sb}\hat{a}^{\dagger}\hat{a}\hat{b}^{\dagger}\hat{b}$, storage self-Kerr (e.g., either $(K_{s}/2)\hat{a}^{\dagger 2}\hat{a}$ when the  coupler is in the off-position, or ancilla-state dependent storage self-Kerr when the coupler is in an on-position), and the storage-coupler dispersive coupling $\chi_{sc}(\Phi_{x,c}(t))\hat{a}^{\dagger}\hat{a}|e_{c}\rangle\langle e_{c}|$. Besides these imperfections, undesired resonances can occur due to side-band transitions of a strong drive or large photon numbers in the storage mode. In the following subsections we provide circuit-quantization-level details of the underlying coupling mechanisms and discuss these imperfections. Moreover, we provide systematic design strategies to suppress these undesired processes. 

\cref{app:tunable_dispersive_coupling_details} provides more details on the tunable dispersive coupling, important considerations for minimizing the storage self-Kerr, and possible undesired resonances when the coupler frequency gets too close to the storage and the ancilla frequencies.  \cref{app:serial_inductance_calculations} focuses on details related to the serial inductances of the buffer modes and how they affect the storage-buffer cross-Kerr and the storage self-Kerr. \cref{app:pump_loss} discusses how the strong sigma flux pump needed for the 3WM process can lead to undesired side-band resonances if the system parameters are not carefully arranged. Lastly, in \cref{app:effective_model_analysis}, we provide more details on the numerical simulation of the cat-qubit bit-flip times based on the effective Hamiltonian of our device. In \cref{app:effective_model_analysis} we also discuss the effective model for two-photon dissipation with added Kerr nonlinearities.

\subsection{Tunable dispersive coupling between a storage mode and an ancilla transmon}
\label{app:tunable_dispersive_coupling_details}

The tunable dispersive interaction between a storage mode and an ancilla transmon is realized by using a tunable-transmon coupler. The Hamiltonian of the storage-coupler-ancilla subsystem is given by
\begin{align}
    \hat{H} &= \omega_{s,0}\hat{a}^{\dagger}\hat{a} 
    \nonumber\\
    &+ 4E_{C,c}\hat{N}_{c}^{2} - E_{J1,c}\cos(\hat{\phi}_{c} + \varphi_{x,c}) - E_{J2,c}\cos\hat{\phi}_{c} 
    \nonumber\\
    &+ 4E_{C,a}\hat{N}_{a}^{2} - E_{J,a}\cos\hat{\phi}_{a} 
    \nonumber\\
    &+ \lambda_{sc}i(\hat{a}^{\dagger}-\hat{a}) \hat{N}_{c} + \lambda_{sa}\hat{N}_{c}\hat{N}_{a} + \lambda_{sa}i(\hat{a}^{\dagger}-\hat{a}) \hat{N}_{a} . 
\end{align}
Here the subscripts ``s'', ``c'', and ``a'' represent the storage, coupler, and the ancilla. Note that ``a'' refers to the ancilla only when it is used as a subscript and it should not be confused with the annihilation operator $\hat{a}$ of the storage. In isolation, the storage mode is a linear quantum harmonic oscillator with a bare frequency $\omega_{s,0}$. The coupler is a tunable-frequency transmon and thus has a SQUID loop with two Josephson junctions with the junction energies $E_{J1,c}$ and $E_{J2,c}$. $\varphi_{x,c} \equiv 2\pi \Phi_{x,c} / \Phi_{0}$ is the dimensionless external flux applied to the coupler's SQUID loop where $\Phi_{0} = h / e$ is the flux quantum. The ancilla is a fixed-frequency transmon with a Josephson junction which has the junction energy $E_{J,a}$. The charging energies of the coupler and the ancilla are given by $E_{C,c}$ and $E_{C,a}$. We assume vanishing offset charge for both the coupler and the ancilla because they are in a ``transmon regime'' and insensitive to the details of the offset charge~\cite{Koch2007_charge_insensitive}. The storage, coupler, and the ancilla qubits are all capacitively coupled with each other through their charge operators $i(\hat{a}^{\dagger}-\hat{a})$, $\hat{N}_{c}$, and $\hat{N}_{a}$, respectively.  

When a coupler mediates tunable coupling between two qubits of the same type (e.g., as in Refs.~\cite{Yan2018Tunable, Sung2021CZ}), the strengths of the qubit-coupler couplings are typically designed symmetrically for the two qubits neighboring the coupler. However, in our system, a coupler mediates tunable interaction between two different types of elements, i.e., a linear storage mode and a nonlinear ancilla transmon. Thus, we design the coupling strengths asymmetrically such that the coupler-storage coupling is much weaker than the coupler-ancilla coupling. Through this asymmetric coupling, we minimize the coupler-induced nonlinearity of the storage mode which is important for implementing a long-lived cat qubit in the storage mode. Note that due to this design choice, the ancilla transmon is subject to strong coupler-induced nonlinearities. However the impact of these induced nonlinearities on the ancilla is minimal given the ancilla's large intrinsic nonlinearity.   

The asymmetry between the coupler-storage coupling and the coupler-ancilla coupling in our system is clearly illustrated by the coupler-induced frequency shifts of the storage and the ancilla shown in \cref{fig:tunable_coupler}(a). As the coupler tunes from its maximum ($8.40\text{GHz}$) to minimum ($5.77\text{GHz}$) frequency, the frequency of the storage is repelled down by only $3~\text{MHz}$ (from $5.349\text{GHz}$ to $5.346\text{GHz}$), whereas the frequency of the ancilla is pushed down by $141~\text{MHz}$ (from $5.211\text{GHz}$ to $5.070\text{GHz}$). Note that the storage frequency is higher than the ancilla frequency and closer to the coupler frequency. Thus, if the coupler-storage and the coupler-ancilla coupling strengths were chosen symmetrically, the storage is expected to hybridize more strongly with the coupler than the ancilla does with the coupler. However, thanks to the large asymmetry in the coupling strengths, the coupler is strongly hybridized only with the ancilla and is weakly hybridized with the storage despite such frequency arrangement.   

Minimizing the induced non-linearities of the storage mode is also crucial for ensuring the dispersive nature of the storage-ancilla interaction. As featured in the last term in \cref{eq:storage_coupler_ancilla_effective_hamiltonian}, the desired dispersive interaction between the storage and the ancilla takes the form $\hat{H}_{sa} = \chi_{sa}\hat{a}^{\dagger}\hat{a}|e_{a}\rangle\langle e_{a}|$ in the dressed eigenbasis. In practice, the storage-ancilla interaction also contains ancilla-state-dependent self-Kerr terms of the storage, i.e.,
\begin{align}
    \hat{H}_{sa} &= \omega_{s}\hat{a}^{\dagger}\hat{a} + \omega_{a}|e_{a}\rangle\langle e_{a}| +  \chi_{sa}\hat{a}^{\dagger}\hat{a}|e_{a}\rangle\langle e_{a}| 
    \nonumber\\
    & \quad + \frac{K_{s, a:|g\rangle}}{2} \hat{a}^{\dagger 2}\hat{a}^{2} |g_{a}\rangle\langle g_{a}|  + \frac{K_{s, a:|e\rangle}}{2} \hat{a}^{\dagger 2}\hat{a}^{2} |e_{a}\rangle\langle e_{a}| . 
\end{align}
Here, $|g_{a}\rangle$, $|e_{a}\rangle$ are the ground and the first excited states of the ancilla. $K_{s, a:|g\rangle}$ and $K_{s, a:|e\rangle}$ are the storage self-Kerr when the ancilla is in the state $|g_{a}\rangle$ and $|e_{a}\rangle$, respectively.  For now, we assume that the coupler always stays in its ground state $|g_{c}\rangle$ and ignore its explicit presence (other than noting all the parameters in the effective Hamiltonian are dependent on the coupler flux $\varphi_{x,c}$). 

Under the above effective Hamiltonian, the storage mode frequencies are given by
\begin{align}
    \omega_{s, a:|g\rangle}^{ (n\rightarrow n+1) } &= E(|n+1, g_{a}\rangle) - E(|n, g_{a}\rangle) 
    \nonumber\\
    &= \omega_{s} + K_{s, a:|g\rangle} n, 
    \nonumber\\
    \omega_{s, a:|e\rangle}^{ (n\rightarrow n+1) } &= E(|n+1, e_{a}\rangle) - E(|n, e_{a}\rangle) 
    \nonumber\\
    &= \omega_{s} + \chi_{sa} + K_{s, a:|e\rangle} n. 
\end{align}
In an ideal storage-ancilla dispersive coupling, the storage mode frequency should be independent of the storage photon number $n$ and is shifted only when the ancilla is excited from $|g_{a}\rangle$ to $|e_{a}\rangle$. However, due to the self-Kerr terms, the storage frequency is shifted not only by the ancilla excitation (by $\chi_{sa}$) but also by its own excitations (by $K_{s, a:|g\rangle} n$ or $K_{s, a:|e\rangle} n$). These unwanted frequency shifts due to self-Kerr terms are particularly detrimental to a cat qubit with a large average photon number. Therefore, the self-Kerr nonlinearities $K_{s, a:|g\rangle}$ or $K_{s, a:|e\rangle}$ should be kept much smaller than the dispersive shift $\chi_{sa}$.  

\begin{figure}[t!]
    \centering
    \includegraphics[width=1.0\columnwidth]{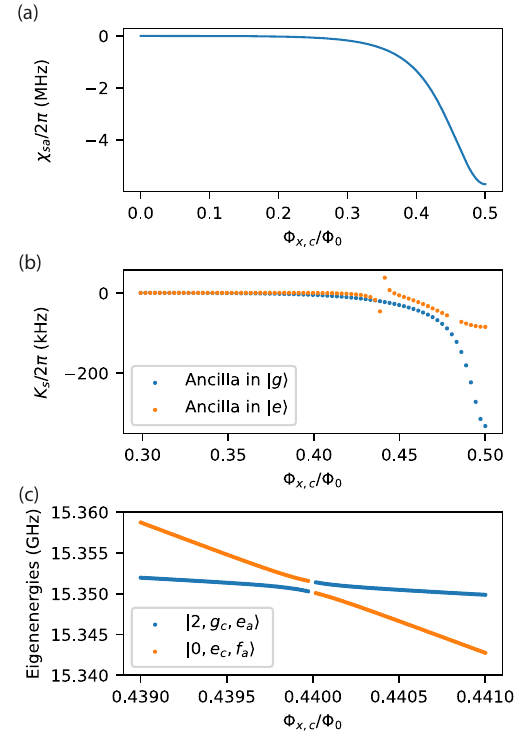}
    \caption{\textbf{Full circuit-quantization-level numerical model of a tunable dispersive interaction.} (a) Storage-ancilla dispersive coupling $\chi_{sa}$ as a function of the external flux of the tunable coupler $\Phi_{x,c}$ in the unit of the flux quantum $\Phi_{0}$. (b) Self-Kerr of the storage mode $K_{s}$ conditioned on the ancilla transmon being in $|g_{a}\rangle$ or $|e_{a}\rangle$ as a function of coupler flux. Note that $K_{s, a:|e\rangle}$ exhibits a resonance feature around $\Phi_{x,c}/\Phi_{0} = 0.44$. (c) Avoided crossing between $|2,g_{c},e_{a}\rangle$ and $|0,e_{c},f_{a}\rangle$ indicating undesired conversion of two storage photons into one coupler excitation ($|g_{c}\rangle\rightarrow |e_{c}\rangle$) and one ancilla excitation ($|e_{a}\rangle\rightarrow |f_{a}\rangle$). This resonance is the root cause of the resonance feature in $K_{s, a:|e\rangle}$ observed above.  }
    \label{app_fig:tunable_coupler_model_predictions}
\end{figure}

In \cref{app_fig:tunable_coupler_model_predictions}(a)--(b), we show the predicted storage-ancilla dispersive shift $\chi_{sa}$ and the storage self-Kerr nonlinearities $K_{s, a:|g\rangle}$ and $K_{s, a:|e\rangle}$ as a function of the coupler flux $\Phi_{x,c} / \Phi_{0} = \varphi_{x,c} / 2\pi$. The parameters of the model used in the prediction are tuned up to reproduce the experimentally-measured mode frequencies $\omega_{s}$, $\omega_{c}$, $\omega_{a}$ and the storage-ancilla dispersive shift $\chi_{sa}$ over the entire range of the coupler flux from $\Phi_{x,c} / \Phi_{0} =0 $ to $\Phi_{x,c} / \Phi_{0} = 0.5$ (shown in \cref{fig:tunable_coupler}(a)--(b)). At the coupler flux $\Phi_{x,c} / \Phi_{0} =0.38$ used for the parity measurement in our experiments, the model predicts $\chi_{sa} / 2\pi = - 0.88 ~\text{MHz}$ (consistent with the experimental data) as well as $K_{s, a:|g\rangle} / 2\pi = -1.7~\text{kHz}$ and $K_{s, a:|e\rangle} / 2\pi = 0.9 ~\text{kHz}$. Thus in this case, $K_{s, a:|g\rangle}$ and $K_{s, a:|e\rangle}$ are approximately three orders of magnitude smaller than $\chi_{sa}$, preserving the dispersive nature of the storage-ancilla interaction even when the storage has, e.g., $10$ photons on average since $|K_{s} \alpha^{2} / \chi_{sa}| \ll 1$ for both $K_{s} = K_{s, a:|g\rangle}$ and $K_{s} = K_{s, a:|e\rangle}$ at $|\alpha|^{2}=10$.  

At the coupler's minimum frequency position ($\Phi_{x,c} / \Phi_{0} =0.5$), the model predicts $\chi_{sa} / 2\pi = - 5.71 ~\text{MHz}$ (to be compared against the experimental value of $-5.73\pm 0.03 ~\text{MHz}$) as well as $K_{s, a:|g\rangle} / 2\pi = -332~\text{kHz}$ and $K_{s, a:|e\rangle} / 2\pi = -84 ~\text{kHz}$. In particular, when the ancilla is in the ground state $|g_{a}\rangle$, the storage self-Kerr $K_{s, a:|g\rangle}$ is only an order-of-magnitude smaller than the dispersive shift $\chi_{sa}$. Thus, in this case, the dispersive nature of the storage-ancilla interaction breaks down in the higher excited state manifold of the storage (e.g., $|K_{s, a:|g\rangle} \alpha^{2} / \chi_{sa}|$ is already close to $1$ with $|\alpha|^{2}=10$ whereas it should be much smaller than $1$). This is why we avoided using the coupler's minimum frequency position as the on-position in our experiments. 

Besides the intolerably large $K_{s} / \chi_{sa}$ ratios, another reason for avoiding $\Phi_{x,c} / \Phi_{0} =0.5$ as the on-position is the undesired resonances that arise as the coupler frequency is lowered and gets close to the storage and ancilla frequencies. For example, the $K_{s, a:|e\rangle}$ curve in \cref{app_fig:tunable_coupler_model_predictions}(b) exhibits a resonance feature around $\Phi_{x,c} / \Phi_{0} =0.44$. The root cause of this feature is the avoided crossing between two energy levels $|2,g_{c},e_{a}\rangle$ and $|0,e_{c},f_{a}\rangle$ (see \cref{app_fig:tunable_coupler_model_predictions}(c)). This transition affects $K_{s, a:|e\rangle}$ because the eigenenergy of the state $|2,g_{c},e_{a}\rangle$ is involved in the self-Kerr $K_{s, a:|e\rangle} \equiv E(|2,g_{c},e_{a}\rangle) - 2E(|1,g_{c},e_{a}\rangle) + E(|0,g_{c},e_{a}\rangle)$. In this $|2,g_{c},e_{a}\rangle\leftrightarrow |0,e_{c},f_{a}\rangle$ transition, which becomes resonant at around $\Phi_{x,c} / \Phi_{0} =0.44$, two storage photons are converted into one coupler excitation ($|g_{c}\rangle \rightarrow |e_{c}\rangle$) and one ancilla excitation ($|e_{a}\rangle \rightarrow |f_{a}\rangle$). As a result, if the on-position coupler flux is placed around or above $\Phi_{x,c} / \Phi_{0} =0.44$, this transition can be brought into resonance and cause undesired coupler and ancilla heating events. Thus, it is crucial to understand and avoid these resonances when realizing a tunable dispersive coupling between a linear mode and a nonlinear qubit.         

\subsection{ATS with serial inductances} \label{app:serial_inductance_calculations}

Here we provide more details of the buffer mode used for stabilizing a cat qubit. To achieve an optimal performance of the cat qubit stabilization (e.g., in terms of bit-flip times), it is important to minimize the self-Kerr of the buffer mode such that the storage mode does not inherit undesired non-linearities from the buffer mode (e.g., storage-buffer cross-Kerr and the storage self-Kerr). In this section, we show that inductances in series with the side junctions of the buffer can have significant impacts on the strength of the self-Kerr of the buffer mode. Moreover, we provide a design strategy to minimize the self-Kerr of the buffer mode on its saddle points by optimizing the Josephson energies of the side junctions.     

As shown in \cref{fig:serial_inductance}, we consider a buffer containining a center inductor (implemented via a ``shallow'' Josephson junction array just with a few, e.g., three junctions in the array), two side junctions, and a shunt capacitor. Importantly, each side junction has a serial inductance. These serial inductances introduce extra flux nodes (with fluxes $\phi_{P_{1}}$ and $\phi_{P_{2}}$) between the ground node (with zero flux by definition) and the buffer node (with flux $\phi$). Thus, the inductive potential of the buffer mode is given by 
\begin{align}
    V &= -NE_{J,\text{array}} \cos\Big{(}\frac{\phi}{N}\Big{)} 
    \nonumber\\
    &\quad -E_{J,1}\cos(\phi + \varphi_{x,1} - \phi_{P_{1}})  + \frac{1}{2}E_{L, P_{1}}\phi_{P_{1}}^{2} 
    \nonumber\\
    &\quad - E_{J,2}\cos(\phi - \varphi_{x,2} - \phi_{P_{2}} ) + \frac{1}{2}E_{L, P_{2}}\phi_{P_{2}}^{2} . 
\end{align}
Here, $E_{J,\text{array}}$ is the junction energy of each junction in the center junction array, $N$ is the number of junctions in the junction array, $E_{J,1},E_{J,2}$ are the junction energies of the two side junctions, and $E_{L,P_{1}},E_{L,P_{2}}$ are the inductive energies of the serial inductances of the side junctions. Moreover $\varphi_{x,1}$ and $\varphi_{x,2}$ are two external fluxes applied to the buffer mode. 

In practice, the serial inductances $L_{P_{1}}$ and $L_{P_{2}}$ are much smaller than the inductances of the Josephson junctions (e.g., by many orders of magnitude), leading to $E_{L,P_{1}} \gg E_{J,1}$ and $E_{L,P_{2}} \gg E_{J,2}$.  In this regime, one may attempt to eliminate the extra fluxes $\phi_{P_{1}}$ and $\phi_{P_{2}}$ by perturbatively optimizing them to minimize the potential~\cite{Quintana2017thesis, Kafri2017}. Then the inductive potential can be simplified into the form (up to an additive constant offset)
\begin{align}
    V &\simeq -NE_{J,\text{array}} \cos\Big{(}\frac{\phi}{N}\Big{)} 
    \nonumber\\
    &\quad  -E_{J,1}\cos(\phi + \varphi_{x,1})  + \frac{E_{J,1}^{2}}{4E_{L, P_{1}}}\cos(2(\phi + \varphi_{x,1})) 
    \nonumber\\
    &\quad  - E_{J,2}\cos(\phi - \varphi_{x,2}) + \frac{E_{J,2}^{2}}{4E_{L, P_{2}}}\cos(2(\phi - \varphi_{x,2})) , 
\end{align}
involving only the buffer flux $\phi$, where we used 
\begin{align}
    &\min_{y\in \mathbb{R}} \Big{(} A\cos(x - y) + \frac{1}{2}By^{2} \Big{)}
    \nonumber\\
    &= \min_{y\in \mathbb{R}} \Big{(} A(\cos(x)\cos(y) + \sin(x)\sin(y)) + \frac{1}{2}By^{2} \Big{)}
    \nonumber\\
    &\simeq \min_{y\in \mathbb{R}} \Big{(} A(\cos(x) + \sin(x)y) + \frac{1}{2}By^{2} \Big{)}
    \nonumber\\
    &= A\cos(x) - \frac{A^{2}}{2B}\sin^{2}(x) 
    \nonumber\\
    &= A\cos(x) + \frac{A^{2}}{4B}\cos(2x) + \text{const} . 
\end{align}
Alternatively, we can get the same relation via (see \cite{Quintana2017thesis})
\begin{align}
    &\min_{y\in \mathbb{R}} \Big{(} A\cos(x - y) + \frac{1}{2}By^{2} \Big{)}
    \nonumber\\
    &= A\Big{(} 1 + \sum_{\nu\ge 1}\frac{2J_{\nu}(\frac{A}{B}\nu)}{\frac{A}{B}\nu^{2}} (\cos(\nu x) - 1)\Big{)}
    \nonumber\\
    &= A\Big{(} 1 + \frac{2J_{1}(\frac{A}{B})}{\frac{A}{B}} (\cos(x) - 1) 
    \nonumber\\
    &\qquad\quad + \frac{2J_{2}(2\frac{A}{B})}{4\frac{A}{B}} (\cos(2x) - 1) + \cdots \Big{)}
    \nonumber\\
    &= A\Big{(} 1 + \frac{\frac{A}{B} + \mathcal{O}((\frac{A}{B})^{3})}{\frac{A}{B}}(\cos(x) - 1) 
    \nonumber\\
    &\qquad\quad + \frac{(\frac{A}{B})^{2} + \mathcal{O}((\frac{A}{B})^{4})}{4\frac{A}{B}} (\cos(2x) - 1) + \cdots \Big{)}
    \nonumber\\
    &\simeq A\cos(x) + \frac{A^{2}}{4B}\cos(2x) + \text{const}. 
\end{align}

\begin{figure*}[t!]
    \centering
    \includegraphics[width=2.0\columnwidth]{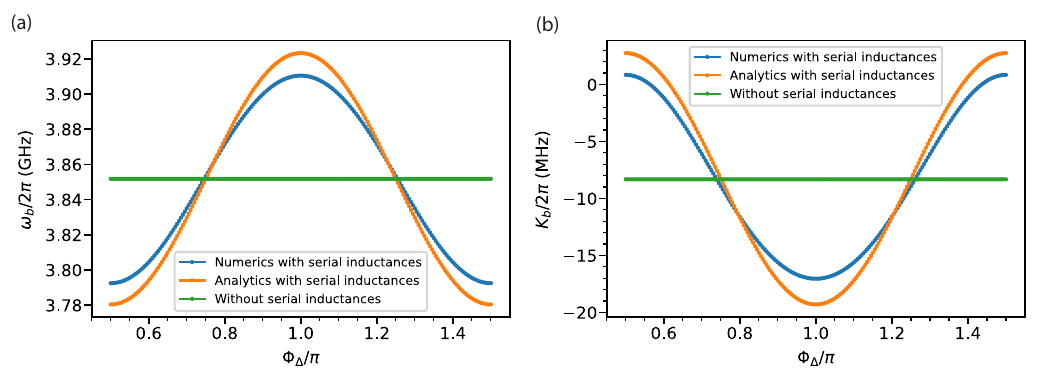}
    \caption{\textbf{Numerical model of a buffer.} (a) Frequency and (b) self-Kerr (i.e., anharmonicity) of the buffer mode in the presence (blue and orange curves) or absence (green curves) of the inductances in series with the side junctions. The blue curves represent predictions via exact numerical circuit quantization and the orange curves represent the analytic expressions in \cref{app_eq:serial_inductance_frequency_and_Kerr_analytic}). Note that the two side junction energies were assumed to be identical such that the analytic expressions are applicable although the numerical approach can be more generally applied. }
    \label{app_fig:serial_inductance_frequency_and_Kerr}
\end{figure*}

To get an intuition on the impact of the serial inductances to the buffer mode, we make simplifying assumptions $E_{J,1} = E_{J,2} = E_{J}$ and $E_{L,P_{1}}=E_{L,P_{2}} = E_{L,P}$ (although the method should be more generally applicable). Then we find 
\begin{align}
    V &\simeq -NE_{J,\text{array}} \cos\Big{(}\frac{\phi}{N}\Big{)} 
    \nonumber\\
    &\quad  -E_{J}\Big{(}\cos(\phi + \varphi_{x,1}) + \cos(\phi - \varphi_{x,2}) \Big{)} 
    \nonumber\\
    &\quad  + \frac{E_{J}^{2}}{4E_{L, P}}\Big{(} \cos(2(\phi + \varphi_{x,1})) + \cos(2(\phi - \varphi_{x,2})) \Big{)} 
    \nonumber\\
    &= -NE_{J,\text{array}} \cos\Big{(}\frac{\phi}{N}\Big{)} -2E_{J}\cos(\varphi_{\Sigma}) \cos(\phi + \varphi_{\Delta})  
    \nonumber\\
    &\quad  + \frac{E_{J}^{2}}{2E_{L, P}}\cos(2\varphi_{\Sigma})   \cos(2(\phi + \varphi_{\Delta}))  , 
\end{align}
where $\varphi_{\Sigma} \equiv (\varphi_{x,1} + \varphi_{x,2}) / 2$ and $\varphi_{\Delta} \equiv (\varphi_{x,1} - \varphi_{x,2}) / 2$. When $\varphi_{\Sigma}$ is an odd integer multiple of $\pi/2$, we have $\cos(2\varphi_{\Sigma}) = -1$ and get 
\begin{align}
    V &\simeq  -NE_{J,\text{array}} \cos\Big{(}\frac{\phi}{N}\Big{)} - \frac{E_{J}^{2}}{2E_{L, P}}  \cos(2(\phi + \varphi_{\Delta}))  . 
\end{align}

If the serial inductances were zero (i.e., $E_{L,P}\rightarrow \infty$), the potential energy is simply given by the one due to the junction array, i.e., $V = -NE_{J,\text{array}} \cos(\frac{\phi}{N})$, independent of the delta flux $\varphi_{\Delta}$. Then in the limit of $N\rightarrow \infty$ (while keeping $E_{J,\text{array}} = NE_{L}$), the potential converges to a harmonic potential $V = \frac{1}{2}E_{L}\phi^{2}$ and the buffer mode's self-Kerr is vanishingly small at all values of $\varphi_{\Delta}$ along the cut defined by $\varphi_{\Sigma} = (2m+1)\pi/2$ for some integer $m \in \mathbb{Z}$. Moreover, the buffer mode will have a constant frequency at all values of $\varphi_{\Delta}$. 

If the serial inductance were not zero (i.e., $E_{L,P}$ is finite), the potential will always have a non-trivial anharmonic contribution $V' = -(E_{J}^{2} / (2E_{L, P}) ) \cos(2(\phi + \varphi_{\Delta}))$. 
To understand the impact of this serial inductance potential term, we add a capacitive charging energy and construct the following Hamiltonian 
\begin{align}
    \hat{H} &= 4E_{C}(\hat{N}-n_{g})^{2} -NE_{J,\text{array}} \cos\Big{(}\frac{\hat{\phi}}{N}\Big{)} 
    \nonumber\\
    &\quad - \frac{E_{J}^{2}}{2E_{L, P}}  \cos(2(\hat{\phi} + \varphi_{\Delta})) . 
\end{align}
Restricting to $n_{g}=0$, and viewing 
\begin{align}
    \hat{H}_{0} = 4E_{C}\hat{N}^{2} + \frac{1}{2}\Big{(}\frac{E_{J,\text{array}}}{N}\Big{)}\hat{\phi}^{2}
\end{align}
as an unperturbed Hamiltonian, we find the correction to the energy eigenvalues to the first order in $\hat{H} - \hat{H}_{0}$. This yields  
\begin{align}
    E_{n} &\simeq  \Big{(}\sqrt{8E_{C}E_{L}} - \frac{E_{C}}{N^{2}} + \frac{2E_{J}^{2}}{E_{L, P}}\sqrt{\frac{2E_{C}}{E_{L}}}\cos(2\varphi_{\Delta}) \Big{)}  n 
    \nonumber\\
    &\quad -\frac{1}{2}E_{C}\Big{(} \frac{1}{N^{2}} + \frac{8E_{J}^{2}}{E_{L, P} E_{L} }\cos(2\varphi_{\Delta}) \Big{)}n(n-1) . 
\end{align}
where we defined $E_{L}\equiv E_{J,\text{array}} / N$. Thus, the frequency and the self-Kerr of the buffer mode are perturbatively given by 
\begin{align}
    \omega_{b} &\simeq \sqrt{8E_{C}E_{L}} - \frac{E_{C}}{N^{2}} + \frac{2E_{J}^{2}}{E_{L, P}}\sqrt{\frac{2E_{C}}{E_{L}}}\cos(2\varphi_{\Delta}) , 
    \nonumber\\
    K_{b} &\simeq -\frac{E_{C}}{N^{2}} - \frac{8E_{C} E_{J}^{2}}{E_{L, P} E_{L} }\cos(2\varphi_{\Delta}) . \label{app_eq:serial_inductance_frequency_and_Kerr_analytic} 
\end{align}

To benchmark the validity of the analytic expressions in \cref{app_eq:serial_inductance_frequency_and_Kerr_analytic}, we perform numerical circuit quantization of the buffer mode in the presence of serial inductors. In the numerics, we do not eliminate the extra modes due to the serial inductors. To make the numerics feasible, it is important to include some small non-zero capacitance in parallel with each serial inductor such that the frequency of the serial inductor mode does not diverge. In practice, we set this capacitance such that the frequency of the serial inductor mode is much higher than that of the buffer mode and confirm that different choices of the capacitance values in this regime do not change the key properties of the buffer mode. As shown in \cref{app_fig:serial_inductance_frequency_and_Kerr}, the analytic expressions in \cref{app_eq:serial_inductance_frequency_and_Kerr_analytic} agree qualitatively with the exact results from numerical circuit quantization. 

Note that on the saddle points (i.e., $\varphi_{\Delta}$ is an odd integer multiple of $\pi/2$ in addition to $\varphi_{\Sigma}$ being also an odd integer multiple of $\pi/2$), we have $\cos(2\varphi_{\Delta}) = -1$ and thus the self-Kerr of the buffer mode is perturbatively given by 
\begin{align}
    K_{b} &\simeq -\frac{E_{C}}{N^{2}} + \frac{8E_{C} E_{J}^{2}}{E_{L, P} E_{L} } . 
\end{align}
Here, the first term is due to the contribution from the junction array and the second term is due to the contribution from the serial inductance. That is, the junction array introduces negative self-Kerr of the buffer mode and the serial inductance introduces positive self-Kerr of the buffer mode at the saddle points. Thus, by balancing the contributions from these two terms, i.e., 
\begin{align}
    \frac{E_{C}}{N^{2}} = \frac{8E_{C} E_{J}^{2}}{E_{L, P} E_{L} }, 
\end{align}
we can minimize the self-Kerr of the buffer mode even when $N$ is finite (e.g., $N=3$ as opposed to $N\rightarrow \infty$). In practice, the side junction energy $E_{J}$ is one of the more flexible design knobs (e.g., not affecting the buffer frequency) among the parameters involved. Hence, choosing it to be $E_{J} = \sqrt{E_{J,\text{array}}E_{L,P}/8N^{3}}$ subject to given values of $E_{J,\text{array}}$ and $E_{L,P}$ is one way to satisfy the above condition for minimizing the buffer self-Kerr. 

\subsection{Buffer-pump-induced parity flip of the storage} 
\label{app:pump_loss}

To realize the desired 3WM interaction ($g_{2}\hat{a}^{2}\hat{b}^{\dagger} + \text{H.c.}$), we pump the sigma flux of the buffer mode~\cite{Lescanne2020}. Here, we discuss how this flux pump can introduce undesired loss mechanism of the storage mode due to driven resonances. In particular, we make use of the Floquet theory in the form of the expanded Hilbert space formalism~\cite{Sambe1973,DiPaolo2022_extensible}. There are simpler alternatives such as naively tabulating all possible frequency-collision conditions. However with this simple approach, one may either accidentally neglect relevant higher-order transitions or become overly conservative by accounting for irrelevant resonance conditions that are prohibited by the ``selection rules''. In contrast, the Floquet approach can serve as a systematic tool for capturing arbitrary higher-order processes as well as estimating the width of the resonance features. 

We are ultimately interested in a composite system consisting of a storage, a flux-pumped buffer, and the additional modes associated with the buffer serial inductances. For now, we focus on an isolated flux-pumped ATS to simplify the presentation. In this case, the Hamiltonian of the buffer is given by 
\begin{align}
    \hat{H}_{b}(t) &= 4E_{C}\hat{N}_{b}^{2} - NE_{J,\text{array}}\cos\Big{(}\frac{\hat{\phi}_{b}}{N}\Big{)} 
    \nonumber\\
    &\quad - (E_{J,1} + E_{J,2})\cos(\varphi_{\Sigma}(t)) \cos(\hat{\phi}_{b} + \varphi_{\Delta}) 
    \nonumber\\
    &\quad + (E_{J,1} - E_{J,2}) \sin(\varphi_{\Sigma}(t))\sin(\hat{\phi}_{b}+\varphi_{\Delta}), 
\end{align}
where $\varphi_{\Sigma}(t)$ is the pumped sigma flux of the buffer and $\varphi_{\Delta}$ is the static delta flux of the buffer. We further assume that the buffer is statically parked at one of its saddle points, e.g., $(\varphi_{\Sigma},\varphi_{\Delta}) = (\pi/2, \pi/2)$. Then, by adding a sigma flux pump with amplitude $\epsilon_{p}$ and frequency $\omega_{p}$, we have $\varphi_{\Sigma}(t) = (\pi/2) + \epsilon_{p}\cos(\omega_{p}t)$ and $\varphi_{\Delta} = \pi/2$ yielding the buffer Hamiltonian
\begin{align}
    \hat{H}_{b}(t) &= 4E_{C}\hat{N}_{b}^{2} - NE_{J,\text{array}}\cos\Big{(}\frac{\hat{\phi}_{b}}{N}\Big{)} 
    \nonumber\\
    &\quad - (E_{J,1} + E_{J,2})\sin( \epsilon_{p}\cos(\omega_{p}t) ) \sin\hat{\phi}_{b} 
    \nonumber\\
    &\quad + (E_{J,1} - E_{J,2}) \cos( \epsilon_{p}\cos(\omega_{p}t) )\cos\hat{\phi}_{b} . 
\end{align}

Note that the driven buffer Hamiltonian above is time-dependent and is periodic in time $t$ with a period of $2\pi / \omega_{p}$. Using the expanded Hilbert space formalism~\cite{Sambe1973,DiPaolo2022_extensible}, we can construct a time-independent Hamiltonian which has the equivalent Floquet spectrum as the time-dependent periodic Hamiltonian. In particular, we introduce a pump mode with a charge operator $\hat{N}_{p}$ and a cosine-phase operator $\cos\hat{\phi}_{p}$ which are defined as 
\begin{align}
    \hat{N}_{p} &= \sum_{n_{p} = -\infty}^{\infty} n_{p}|n_{p}\rangle\langle n_{p}| , 
    \nonumber\\
    \cos\hat{\phi}_{p} &= \frac{1}{2}\sum_{n_{p} = -\infty}^{\infty} ( |n_{p}+1\rangle\langle n_{p}| + \text{H.c.} ). \label{app_eq:pump_mode_operator_definitions}
\end{align}
Here, one may interpret $n_{p}$ as the number of pump photons in the pump mode. Then, the equivalent time-independent Floquet Hamiltonian is given by 
\begin{align}
    \hat{H}_{F} &= \hat{H}_{b} + \hat{H}_{p} + \hat{H}_{bp}, 
\end{align}
where  $\hat{H}_{b} = 4E_{C}\hat{N}_{b}^{2} - NE_{J,\text{array}}\cos(\frac{\hat{\phi}_{b}}{N}) + (E_{J,1} - E_{J,2}) \cos\hat{\phi}_{b}$ is the static buffer Hamiltonian, $\hat{H}_{p} = \omega_{p}\hat{N}_{p}$ is the Hamiltonian of the pump mode, and
\begin{align}
    \hat{H}_{bp} &= - (E_{J,1} + E_{J,2})\sin( \epsilon_{p}\cos\hat{\phi}_{p} ) \sin\hat{\phi}_{b} 
    \nonumber\\
    &\quad - (E_{J,1} - E_{J,2})  ( 1 - \cos( \epsilon_{p}\cos\hat{\phi}_{p} ) ) \cos\hat{\phi}_{b} . 
\end{align}
is the interaction Hamiltonian between the buffer mode and the pump mode due to the sigma flux pump.  

From the pump-mode Hamiltonian $\hat{H}_{p} = \omega_{p}\hat{N}_{p}$, we see that the pump mode releases an excess energy of $\omega_{p}$ (i.e., pump frequency) when it loses one pump photon. Such an excess energy is gained by system through the interaction Hamiltonian $\hat{H}_{bp}$ to either realize a desired driven process or induce an undesired transition. For illustration, we expand the interaction Hamiltonian to the first order in the pump amplitude $\epsilon_{p}$, i.e., 
\begin{align}
    \hat{H}_{bp} &= - (E_{J,1} + E_{J,2}) \epsilon_{p}\cos\hat{\phi}_{p} \sin\hat{\phi}_{b} + \mathcal{O}(\epsilon_{p}^{2}) . 
\end{align}
The term $\cos\hat{\phi}_{p}$ acts on the pump mode and either adds or takes away one pump photon in the pump mode (see \cref{app_eq:pump_mode_operator_definitions}). The term $\sin\hat{\phi}_{b}$ acts on the bare buffer mode. For an isolated buffer mode, this term can induce various transitions in the buffer mode which involve an odd number of excitation gain or loss (e.g., $|0\rangle\leftrightarrow |1\rangle$ and $|0\rangle\leftrightarrow |3\rangle$) due to the parity. When the resonant frequency of one of these transitions, say $|0\rangle\leftrightarrow |3\rangle$, coincides with the frequency of the pump mode, i.e., $\omega_{p} = \omega_{b, 0\leftrightarrow 3}$, an avoided crossing happens in the expanded Hilbert space between the two relevant states $|0, n_{p}=0\rangle \leftrightarrow |3, n_{p}=-1\rangle$. This avoided crossing is then translated into a time-dependent $|0\rangle\leftrightarrow |3\rangle$ transition in the original Hilbert space of the system after the pump mode is projected out~\cite{DiPaolo2022_extensible}.       

In our system, the buffer mode is not isolated and is instead coupled to the storage mode as well as the additional modes associated with the serial inductances. Even then, most of the above analysis is still applicable with the only difference that the term $\sin\hat{\phi}_{b}$ can now introduce more complex transitions involving the modes that are hybridized with the buffer mode. The specific transition most relevant to the cat qubit implementation is a three-wave mixing (3WM) process $\hat{a}^{2}\hat{b}^{\dagger} + \text{H.c.}$ on the dressed modes of the storage and the buffer. In particular, this transition is realized by the matrix elements of the form $\langle \widetilde{n-2,1} | \sin\hat{\phi}_{b} | \widetilde{n,0} \rangle$ and $\langle \widetilde{n,0} | \sin\hat{\phi}_{b} | \widetilde{n-2,1} \rangle$, where $|\widetilde{n,m}\rangle$ is a static dressed eigenstate containing $n$ storage photons and $m$ buffer photons in the absence of the pump (here we assume that the additional modes associated with the serial inductances always stay in their ground states given the high frequency of these modes well above $10\text{GHz}$). Let $\bar{\omega}_{s}$ and $\bar{\omega}_{b}$ denote the Stark-shifted frequencies of the storage and the buffer under the buffer pump. Then when the resonance condition $\omega_{p} = 2\bar{\omega}_{s} - \bar{\omega}_{b}$ is met, an avoided crossing between the two Floquet dressed states $|\overline{n,0,n_{p}=-1}\rangle$ and $|\overline{n-2,1,n_{p}=0}\rangle$ occurs in the expanded Hilbert space, leading to the desired 3WM process in the system when the pump mode is projected out.

\begin{figure}[t!]
    \centering
    \includegraphics[width=\columnwidth]{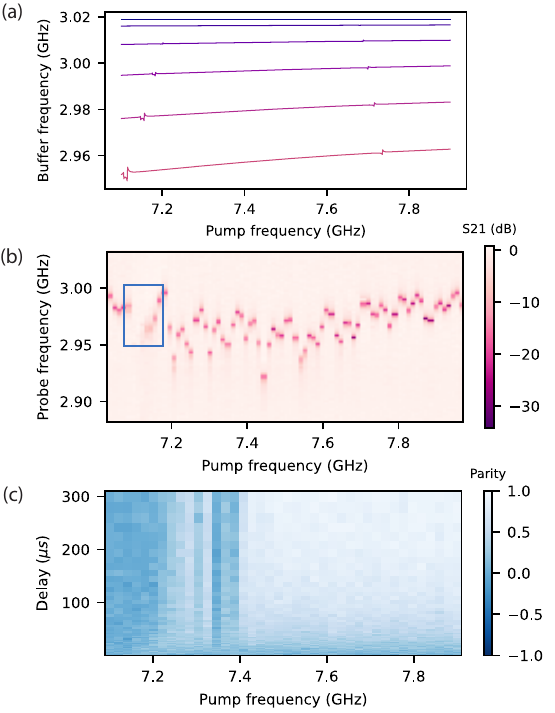}
    \caption{\textbf{Pump induced parity changes.} (a) Numerically simulated buffer frequency as a function of the pump frequency for different pump amplitudes $\epsilon_{p} \in [0, 0.02, 0.04, 0.06, 0.08, 0.1]$. As the pump amplitudes increase the buffer frequency is shifted lower due to the buffer stark shift.  In addition to the desired 3WM condition near $7.7~\text{GHz}$ we observe an additional resonance near $7.2~\text{GHz}$ corresponding to $(3\omega_b+\omega_a)/2$.  (b) Experimentally measured spectrum of the buffer mode resonance as a function of pump frequency for $\epsilon_p \sim 0.08-0.1$. There is a notable loss in contrast in the buffer spectrum in a frequency band below $7.2~\text{GHz}$, as indicated by the blue rectangle.  Note the 3WM condition (expected at $\sim 7.7~\text{GHz}$) is not observable on a scan as coarse as this. (c) Parity of the storage mode as a function of time starting in a superposition of $|0\rangle/|1\rangle$ Fock states prepared with $3~\mathrm{\mu s}$ of two-photon dissipation.  The storage mode is expected to decay to the vacuum state with parity of $+1$, but we observe multiple frequency bands where the steady state parity of the storage mode is far from $+1$. }
    \label{fig:pump_loss}
\end{figure} 

The above expanded Hilbert space framework allows us to systematically search for any undesired resonances under the buffer pump. That is, one can diagonalize the time-independent Floquet Hamiltonian $\hat{H}_{F}$, including the pump mode, and then look for undesired avoided crossings between a pair of Floquet dressed states. Since the buffer mode is heavily attenuated with a large decay rate $\kappa_{b}$, we primarily focus on the spurious transitions involving the buffer ground state. In particular, we compute the Stark-shifted frequency of the buffer mode (i.e., $\bar{\omega}_{b} \equiv E(|\overline{0,1,n_{p}=0}\rangle) - E(|\overline{0,0,n_{p}=0}\rangle)$) which are shown in \cref{fig:pump_loss}(a) for various pump amplitudes of $\epsilon_{p} \in [0, 0.02, 0.04, 0.06, 0.08, 0.1]$. The underlying parameters of the model Hamiltonian are tuned up to match the experimentally observed energy spectrum (e.g., as in \cref{fig:serial_inductance}(b)). 

As the pump amplitude $\epsilon_{p}$ is increased from $0$ (dark purple) to $0.1$ (light purple) in \cref{fig:pump_loss}(a), the frequency of the buffer mode is negatively Stark shifted. Note that there are kinks in the buffer frequencies both around $\omega_{p} / 2\pi \sim 7.7~\text{GHz}$ and $\omega_{p} / 2\pi \sim 7.2~\text{GHz}$. We identify that the former feature around $\omega_{p} / 2\pi \sim 7.7~\text{GHz}$ is due to the desired 3WM avoided crossing between $|\overline{2,0,n_{p}=-1}\rangle$ and $|\overline{0,1,n_{p}=0}\rangle$ (i.e., $\hat{a}^{2}\hat{b}^{\dagger} + \text{H.c.}$) with the resonance condition of $\omega_{p} = 2\bar{\omega}_{s} - \bar{\omega}_{b}$. On the other hand, the latter feature around $\omega_{p} / 2\pi \sim 7.2~\text{GHz}$ is due to an undesired avoided crossing between $|\overline{1,3,n_{p}=-2}\rangle$ and $|\overline{0,0,n_{p}=0}\rangle$ (i.e., $\hat{a}^{\dagger}\hat{b}^{\dagger 3} + \text{H.c.}$) with the resonance condition of $2\omega_{p} = \bar{\omega}_{s} +3\bar{\omega}_{b}$. Upon adiabatic elimination of the buffer mode, this process leads to a pump-induced heating of the storage mode. Note that this undesired resonance is a second-order process, involving two pump photons, that could have been overlooked if one only looked for first-order processes manually without the help of a systematic Floquet analysis.  

\cref{fig:pump_loss}(b) and (c) show various experimental signatures of the undesired resonant process $\hat{a}^{\dagger}\hat{b}^{\dagger 3} + \text{H.c.}$ observed in our device. In \cref{fig:pump_loss}(b) we show the measured buffer frequency versus buffer-pump-frequency, similar to that simulated in \cref{fig:pump_loss}(a). The variation in the buffer mode frequency versus the buffer-pump frequency is due to variations in flux pump power delivery with frequency which makes direct observation of resonances difficult. Notably though, below approximately $\omega_{p} / 2\pi \sim 7.2\text{GHz}$ we observe a reduction in the resonant buffer transmission contrast. In \cref{fig:pump_loss}(c) we show the corresponding measured photon number parity of the storage mode as a function of time after preparation in the $|0\rangle/|1\rangle$ Fock state, under the same buffer pump conditions. In the absence of a nonlinear resonance or for the desired 3WM process, the storage mode would decay into the ground state of party $+1$. Here we see the impact of the undesired resonance in a more pronounced manner, where in a broad frequency band around $\omega_{p} / 2\pi \sim 7.2\text{GHz}$, the storage mode parity does not decay to $+1$, but rather reaches a steady-state parity far from $+1$. This is consistent with the action of buffer-pump-induced heating due to the undesired $\hat{a}^{\dagger}\hat{b}^{\dagger 3} + \text{H.c.}$ resonant process.   

If the two resonant pump frequencies $\omega_{p} = 2\bar{\omega}_{s} - \bar{\omega}_{b}$ (desired 3WM) and $\omega_{p} = (\bar{\omega}_{s} +3\bar{\omega}_{b})/2$ (undesired resonance) are close to each other, the phase coherence of the storage mode at the desired 3WM condition is degraded by the near-resonant pump-induced storage heating. In our system, we have carefully chosen the device parameters to ensure that the desired 3WM condition is far-detuned from the undesired resonance condition by about $500~\text{MHz}$ in pump frequency. Thus, the storage-mode coherence times in our device are not significantly degraded by the two-photon dissipation as demonstrated in \cref{fig:storage_coherence}(d),  \cref{fig:storage_coherence}(e), and \cref{fig:cat_lifetimes}(b).  

We lastly remark that there is another resonance feature in \cref{fig:pump_loss}(c) around $\omega_{p} / 2\pi \sim 7.35\text{GHz}$ (i.e., degradation in the storage parity) that is not captured by our model prediction in \cref{fig:pump_loss}(a). In this regard, we note that our model does not account for the decay of the buffer mode. If the buffer loss is accounted for in the model, we expect that each resonance feature will be broadened in the pump frequency due to the decay-induced linewidth of the buffer mode. Specifically in our system, the buffer loss is engineered with a metamaterial bandpass filter with a wide passband and multiple filter modes. Thus we anticipate that the pump-induced resonance features in our system are not only broadened by the loss channel but also come with additional ripples associated with internal modes of the multi-pole metamaterial filter. The latter could be a root cause of the unexplained parity degradation at the pump frequency of $\omega_{p} / 2\pi \sim 7.35\text{GHz}$. We leave a more detailed analysis of this phenomena as a future research direction for engineering driven-dissipative systems.     

\subsection{Analysis of cat-qubit bit-flip times with an effective model}
\label{app:effective_model_analysis}

In our cat-qubit experiments, the coupler is parked in the off-position for the bulk of a control sequence, and is pulsed to the on-position only in the final measurement step (or sometimes also in the state-preparation step). Thus, to understand the bit-flip times, which are not sensitive to the state-preparation and measurement errors, it suffices to consider an effective Hamiltonian with the coupler in the off-position. Hence we consider the effective Hamiltonian in \cref{eq:simple_two_photon_main_text} which is duplicated here for the readers' convenience. 
\begin{align}
    \hat{H} &= \Big{(}g_{2}(\hat{a}^{2}-\alpha^{2})\hat{b}^{\dagger} + \text{H.c.}\Big{)}  + \chi_{sb}\hat{a}^{\dagger}\hat{a}\hat{b}^{\dagger}\hat{b} + \frac{K_{s}}{2}\hat{a}^{\dagger 2}\hat{a}^{2} . 
    \label{app_eq:two_photon_dissipation_effective_hamiltonian}
\end{align}
In addition to this effective Hamiltonian, we further consider a Lindblad dissipator of the form 
\begin{align}
    \mathcal{L}(\hat{\rho}(t)) &= \kappa_{b}\mathcal{D}[\hat{b}]\hat{\rho}(t) + \kappa_{1}\mathcal{D}[\hat{a}]\hat{\rho}(t) + \kappa_{\phi}\mathcal{D}[\hat{a}^{\dagger}\hat{a}]\hat{\rho}(t). 
    \label{app_eq:two_photon_dissipation_lindblad_terms}
\end{align}

To understand how the storage-buffer cross-Kerr and the storage self-Kerr degrade the bit-flip performance of a cat qubit, we for now ignore the storage decay and dephasing times, i.e., $\kappa_{1} = \kappa_{\phi}=0$ (in the numerical simulations below, we consider non-zero decay and dephasing rates). Then we apply the effective operator formalism~\cite{ReiterEffective} to understand the system analytically. We consider the subspace with the buffer mode in its ground state as the ground-state manifold and the rest as the excited-state manifold. Moreover, we ignore the second and higher excited states of the buffer mode in our analytic calculations.  

Adopting the notation in Ref.~\cite{ReiterEffective}, we identify $\hat{L}$, $\hat{H}_{e}$ as $\hat{L} = \sqrt{\kappa_{b}}$, $\hat{H}_{e} = \chi_{sb}\hat{a}^{\dagger}\hat{a}$ and find that the non-Hermitian Hamiltonian of the excited-state manifold is given by $\hat{H}_{\text{NH}} = \hat{H}_{e} - (i/2)\hat{L}^{\dagger}\hat{L}  = \chi_{sb}\hat{a}^\dagger \hat{a} -i \kappa_b/2$. Then in the limit of $\chi_{sb}\langle \hat{a}^\dagger \hat{a}\rangle/\kappa_b \ll 1$, we have 
\begin{align}
    \hat{H}_{\text{NH}}^{-1} = \frac{2i}{\kappa_b}\Big{(}1-\frac{2i\chi_{sb}}{\kappa_b}\hat{a}^\dagger \hat{a} + O((\frac{\chi_{sb}}{\kappa_b}\hat{a}^\dagger \hat{a})^2)\Big{)} . 
\end{align} 
We further identify $\hat{V}_{+}$, $\hat{H}_{g}$ as $\hat{V}_{+} = g_{2}(\hat{a}^{2} - \alpha^{2}) = \hat{V}_{-}^{\dagger}$, $\hat{H}_{g} = (K_{s}/2)\hat{a}^{\dagger 2}\hat{a}^{2}$ and arrive at the effective master equation 
\begin{align}
    \frac{d\hat{\rho}}{dt}=-i[\hat{H}_{\text{eff}},\hat{\rho}]+\frac{4g_2^2}{\kappa_b}D[\Big{(}1-\frac{2i\chi_{sb}}{\kappa_b}\hat{a}^\dagger \hat{a}\Big{)}(\hat{a}^2-\alpha^2)]\hat{\rho}, \label{app_eq:effective_master_equation_overview}
\end{align}
with
\begin{align}
    \hat{H}_{\text{eff}}=-\frac{2 g_2^2\chi_{sb}}{\kappa_b^2}(\hat{a}^{\dagger 2}-\alpha^2)\hat{a}^\dagger \hat{a}(\hat{a}^2-\alpha^2) + \frac{K_{s}}{2} \hat{a}^{\dagger 2} \hat{a}^2 ,  
    \label{app_eq:effective_master_equation_hamiltonian}
\end{align}
by using $\hat{L}_{\text{eff}} = \hat{L}\hat{H}_{\text{NH}}^{-1}\hat{V}_{+}$ and $\hat{H}_{\text{eff}} = -\hat{V}_{-} ( \hat{H}_{\text{NH}}^{-1} + (\hat{H}_{\text{NH}}^{-1})^{\dagger} ) \hat{V}_{+} /2  + \hat{H}_{g}$. 

To get an intuition on this effective master equation, we first consider a special case where the storage-buffer cross-Kerr is absent, i.e., $\chi_{sb}=0$. In this case we have two-photon dissipation $d\hat{\rho} / dt=-i[\hat{H}_{\text{eff}},\hat{\rho}]+\kappa_2 D[\hat{a}^2-\alpha^2]\hat{\rho}$ with an added storage Kerr nonlinearity $\hat{H}_{\text{eff}}= (K_{s}/2) \hat{a}^{\dagger 2} \hat{a}^2$. The addition of the storage mode Kerr nonlinearity ($K_{s}\ll\kappa_2$) to the dissipatively stabilized cat qubit has a relatively harmless effect in the case without the storage-buffer cross-Kerr. In particular, the two-photon drive can simply be reapportioned between dissipative and Hamiltonian terms to yield a hybrid dissipative-Kerr cat stabilization~\cite{Puri2019Stabilized}, i.e., 
\begin{align}
    \frac{d\hat{\rho}}{dt} &= -i\Big{[}\frac{K_{s}}{2}(\hat{a}^{\dagger 2}-\alpha'^{* 2})(\hat{a}^{\dagger 2}-\alpha'^{2}) ,\hat{\rho}\Big{]}
    \nonumber\\
    &\quad +\frac{4g_2^2}{\kappa_b}D[\hat{a}^2-\alpha'^2]\hat{\rho} , 
\end{align} 
where the steady-state coherent-state amplitude squared is now  $\alpha'^2=\alpha^2/(1+ iK_{s} / \kappa_2)$ with $\kappa_{2} = 4g_{2}^{2} / \kappa_{b}$. This indicates that the storage Kerr nonlinearity is simply used for a Kerr-cat type stabilization making use of some of the two-photon drive from the dissipative stabilization. This causes a slight change in amplitude and the phase of the stabilized cat qubit but does not degrade its bit-flip times. Thus, the storage self-Kerr alone without the storage-buffer cross-Kerr is not detrimental to a dissipative cat qubit. 

As discussed in detail in the main text, a non-zero storage-buffer cross-Kerr causes dephasing of the storage mode when the buffer is excited. For example, such dephasing manifests as a distorted engineered jump operator $(1- 2i\chi_{sb}\hat{a}^\dagger \hat{a} / \kappa_b)(\hat{a}^2-\alpha^2)$. This additional dephasing can then significantly degrade the bit-flip times of a cat qubit when combined with various other mechanisms such as storage loss, dephasing, and self-Kerr which bring the storage mode outside of the cat qubit manifold and consequently excite the buffer mode. 

\begin{figure}[t!]
    \centering
    \includegraphics[width=\columnwidth]{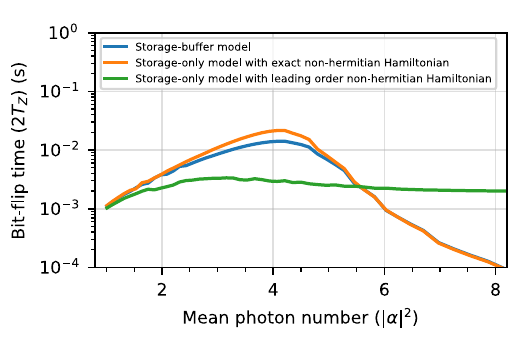}
    \caption{\textbf{Comparison of effective models for two-photon dissipation with added Kerr nonlinearities.} We simulate the case where $E_{J}$ and $E_{L,P}$ are double the experimental values (as in the brown curve of \cref{fig:cat_lifetimes}(a)).  In this regime the effect of the storage-buffer cross-Kerr nonlinearities are significant. We compare a full model which explicitly includes both the storage mode and the buffer mode (blue), an effective model with the buffer adiabatically eliminated with an exact expression of $H_{\text{NH}}^{-1}$ (orange), and an effective model with the buffer adiabatically eliminated which uses only the leading-order term in $H_{\text{NH}}^{-1}$ (green).}
    \label{app_fig:effective_model_comparison}
\end{figure}

Note that the effective master equation in \cref{app_eq:effective_master_equation_overview,app_eq:effective_master_equation_hamiltonian} is derived only up to the first order in $\chi_{sb}\hat{a}^{\dagger}\hat{a} / \kappa_{b}$ using a series-expanded expression for $\hat{H}_{\text{NH}}^{-1}$. While these simple expressions give a useful intuition on how the storage-buffer cross-Kerr degrades the bit-flip times of a cat qubit, as shown in \cref{app_fig:effective_model_comparison}, the resulting effective dynamics (green curve) does not agree quantitatively with the exact dynamics given by the full storage-buffer model (blue curve), especially when $|\chi_{sb}\alpha^{2}|$ is comparable to $\kappa_{b}$. However, one can drastically improve the accuracy of the effective master equation by not using a series expansion for $\hat{H}_{\text{NH}}^{-1}$, and instead using an exact expression (orange curve): 
\begin{align}
    \hat{H}_{\text{NH}}^{-1} = \sum_{n=0}^{\infty} \frac{ 2i }{  \kappa_{b} + 2i\chi_{sb} n } |n\rangle\langle n| . 
    \label{app_eq:exact_non_hermitian_hamiltonian_inverse} 
\end{align}
The computational cost of solving the effective master equation does not increase with this modification while the accuracy is substantially improved. 

The numerical simulations in \cref{app_fig:effective_model_comparison} clearly illustrate the damaging effects of the storage-buffer cross-Kerr. In our numerical simulations, we use the measured parameters when applicable and otherwise use the parameters predicted from circuit-quantization-level models. For the buffer decay rate, we use $\kappa_{b}/2\pi = 10.7~\text{MHz}$. For $\kappa_{1}$ and $\kappa_{\phi}$, we use the relations 
\begin{align}
    \frac{1}{T_{1}} &= \kappa_{1},\quad \frac{1}{T_{2}} = \frac{1}{2}\kappa_{1} + \frac{1}{2}\kappa_{\phi}, 
\end{align}
and estimate $\kappa_{1}$ and $\kappa_{\phi}$ based on the measured storage-mode $T_{1}$ and $T_{2}$ coherence times under the two-photon dissipation (i.e., $T_{1} = 79~\mathrm{\mu s}$ and $T_{2} = 116~\mathrm{\mu s}$). 

For the storage self-Kerr $K_{s}$ and the storage-buffer cross-Kerr $\chi_{sb}$, we use the values predicted by a detailed modeling of our storage-buffer system, accounting for the presence of serial inductances in the buffer mode (see \cref{sec:storage_linearity}). In particular, when we use the circuit parameters that are tuned up to reproduce the experimentally measured buffer frequency spectrum, we predict $K_{s}/2\pi = 1.1~\text{kHz}$ and $\chi_{sb} / 2\pi = 156~\text{kHz}$. These $K_{s}$ and $\chi_{sb}$ values are used in the dashed turquoise curve in \cref{fig:cat_lifetimes}(a). When we use the hypothetical circuit parameters where the average buffer side junction energy $E_{J}$ and the serial inductance $L_{P}$ are doubled compared to the actual device parameters, we predict $K_{s}/2\pi = 9.8~\text{kHz}$ and $\chi_{sb} / 2\pi = 2029~\text{kHz}$ which are used in the dashed brown curve in \cref{fig:cat_lifetimes}(a). 

Since the cat qubit bit-flip times are not sensitive to SPAM errors, we do not explicitly simulate the realistic readout of the storage mode using an ancilla transmon. Instead, we assume ideal state preparation and measurement of a cat qubit in the $|0\rangle/|1\rangle$ basis (i.e., Z basis). Then, we extract the bit-flip time using the same approach used in the experiments, including the convention of the bit-flip time as defined in \cref{app:error_rate_convention}. 

From \cref{fig:cat_lifetimes}(a), it is worth noting that experimentally measured bit-flip times exceed the simulated bit-flip times (dashed turquoise curve) by a multiplicative offset in the regime where the bit-flip error rates are exponentially suppressed in $|\alpha|^{2}$. One of the possible reasons behind this slight discrepancy is that we consider white-noise pure dephasing of the storage mode in our effective model above. On the other hand, the pure storage mode dephasing in our device may have a colored-noise spectrum (e.g., more concentrated on the lower frequencies) as opposed to a frequency-independent white-noise spectrum. However, one can phenomenologically tune the white-noise dephasing rate $\kappa_{\phi}$ to attempt to explain the observed cat-qubit bit-flip times. In the dashed blue curve in \cref{fig:cat_lifetimes}(a) and in all pulsed cat-qubit stabilization simulations in \cref{fig:pulsed_stabilization}(b), we have indeed empirically reduced $\kappa_{\phi}$ by a factor of $2$ such that the simulations agree better with the static bit-flip times measured experimentally. However, we have otherwise kept all other parameters the same.

\section{$g_{2}$ and $\kappa_2$ fitting} 
\label{app:g2_fitting}

\begin{figure}[t!]
    \centering
    \includegraphics[width=\columnwidth]{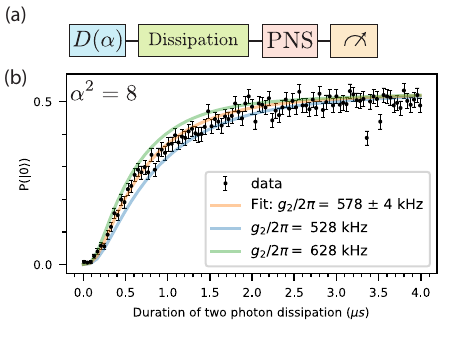}
    \caption{\textbf{$g_2$ fitting.} (a) Pulse sequence for the experiment.  We displace the storage mode into a coherent state.  We apply a pure two-photon dissipation without the buffer drive to map the storage mode into the $|\hat{n}=0\rangle$/$|\hat{n}=1\rangle$ manifold.  We measure the storage mode vacuum population using a number selective $\pi$-pulse on the ancilla.  (b) Experimental data of the vacuum population of the storage mode as a function of the two-photon dissipation duration starting from $|\alpha|^2=8$. Curves correspond to simulations of the experiment with an effective model containing the storage mode and buffer mode. The model fit (orange) yields $g_2/2\pi= 578\pm 4 \mathrm{kHz}$, from which we infer $\kappa_{2}/2\pi = 124\pm 2~\text{kHz}$ using the independently measured value of $\kappa_{b}/2\pi = 10.7~\text{MHz}$.  Reference curves (blue, green) are also included to show the sensitivity of the convergence.}
    \label{app_fig:g2_fitting}
\end{figure} 
To determine $g_{2}$ and $\kappa_2$ we fit the decay of the storage mode to the vacuum state from a coherent state when a pure two-photon dissipation $\kappa_{2}\mathcal{D}[\hat{a}^{2}]$ is applied.  The pure two-photon dissipation maps the storage mode to the $|\hat{n}=0\rangle$/$|\hat{n}=1\rangle$ manifold.  Since we start from an initial state with  vanishing average parity this results in the vacuum population of the storage mode quickly approaching approximately $0.5$ where the convergence time scale is determined by the two-photon dissipation strength. We fit this vacuum population curve to determine $g_{2}$ and $\kappa_2$.  

The sequence used for the experiment is shown in \cref{app_fig:g2_fitting}(a).  The storage mode is displaced to a coherent state $|\alpha\rangle$ with $|\alpha|^{2}=8$, two-photon dissipation is applied for variable durations, and finally the vacuum population of the storage mode is read out using a photon number selective pulse on the ancilla. \cref{app_fig:g2_fitting}(b) shows the vacuum population of the storage mode as a function of the two-photon dissipation duration. At the beginning the vacuum population of the storage mode is negligible because the initial coherent state has low overlap with it.  On the timescale of $3~\mathrm{\mu s}$ the storage mode has mostly converged to the $|\hat{n}=0\rangle$/$|\hat{n}=1\rangle$ manifold as indicated by the vacuum population reaching approximately $0.5$.  The rest of the population in $|\hat{n}=1\rangle$ will decay to the vacuum state due to the storage $T_1$. By fitting the experimental data against a Lindblad master equation model, we extract $g_{2}/2\pi = (\kappa_2 \kappa_b)^{1/2}/4\pi = 578\pm 4~\text{kHz}$. From this value of $g_{2}/ 2\pi$, and the independently measured value of $\kappa_{b}/2\pi=10.7$~MHz, we infer a value of $\kappa_{2}/2\pi= 124 \pm 2~\text{kHz}$ ($\kappa_{2} \approx 4g_{2}^{2} /\kappa_{b}$ when the buffer mode is adiabatically eliminated).

\section{Storage conditional phase measurement} \label{app:storage_conditional_phase}

\begin{figure}[t!]
    \centering
    \includegraphics[width=\columnwidth]{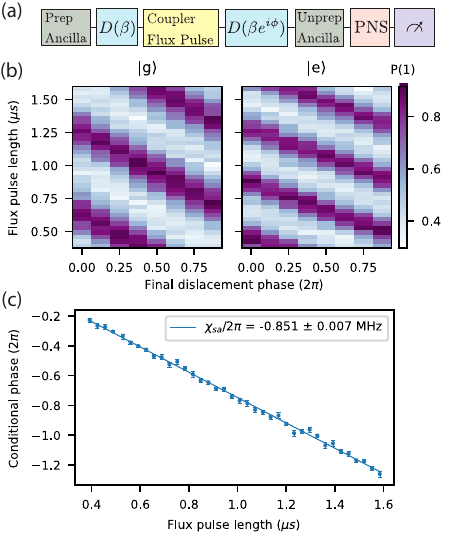}
    \caption{\textbf{Storage conditional phase measurement.} (a) A high-level description of the pulse sequence for characterizing the storage-ancilla dispersive coupling strength. We prepare the ancilla into either $|g\rangle$ or $|e\rangle$, displace the storage, apply a coupler flux pulse with a variable pulse length, displace back the storage with a variable phase, unprepare the ancilla state (inverse of the state preparation), and finally measure the storage mode vacuum population using a vacuum-selective pulse on the ancilla. (b) Fringes measured for two different ancilla states $|g\rangle$ and $|e\rangle$ as a function of the flux pulse length.  From these fringes we determine the storage conditional phase for each flux pulse length. (c) Storage conditional phase as a function of the flux pulse length. From a fitted linear relationship between the flux pulse length and the storage conditional phase, we extract $\chi_{sa} / 2\pi = -0.851\pm 0.007~\text{MHz}$. This data set was collected at the ``parity measurement'' position of the coupler.}
    \label{app_fig:conditional_phase}
\end{figure} 
We use a storage conditional phase measurement to determine the storage-ancilla dispersive coupling and the flux pulse length needed for parity measurement.  The pulse sequence for the conditional phase measurement is shown in \cref{app_fig:conditional_phase}(a).  The conditional phase measurement involves preparing the storage into a small coherent state, preparing the ancilla state ($|i\rangle$ where $i=g,e$), applying a flux pulse with a variable pulse length $t$, unpreparing the ancilla state, displacing the storage mode with a variable phase (8 phases $\phi$ from $0$ to $7\pi/8$ in increments of $\pi/8$), and finally reading out the storage mode vacuum population. Both the initial and final displacement use an amplitude of $\alpha = 0.5$.  The raw data resulting from this experiment is shown in \cref{app_fig:conditional_phase}(b).  For each ancilla state we observe fringes indicating the evolving phase of the storage mode.  

During the course of the flux pulse the initial storage mode coherent state $|\alpha\rangle$ approximately evolves into a final coherent state with the same amplitude but a different phase $|\alpha e^{i\phi_0(t, i)}\rangle$.  We aim to extract $\phi_0(t, i)$ from the fringes of the conditional phase measurement for each ancilla state and flux pulse length.  We determine $\phi_0(t, i)$ by fitting the measured vacuum population to the model
\begin{align}
    P_0 = A e^{-\alpha^2 (2-2\cos{(2\pi\phi-2\pi\phi_0)})} + C
\end{align}
In \cref{app_fig:conditional_phase}(c) we plot the difference in storage mode phase between the ancilla being in $|g\rangle$ versus $|e\rangle$ (i.e. $\phi_0(t, e)-\phi_0(t, g)$).  By fitting the phase difference to a linear model in pulse length $t$, we determine the dispersive shift $\chi_{sa}$ from the ancilla on the storage mode. The flux pulse length required to achieve a $\pi$ conditional phase is not exactly $\pi/|\chi_{sa}|$ because of transients of the flux pulse.  To determine the parity measurement flux pulse length we find where the storage conditional phase reaches $-\pi$.  Note that in addition to determining the dispersive coupling strength and flux pulse length for the parity measurement, we determine the frequency of the storage with the coupler in the on-position by fitting the storage phase as a function of the flux pulse length with the ancilla in $|g\rangle$.  

For the storage vacuum population measurements, a weak number-selective pulse is applied on the ancilla. The number selective pulses we use have a duration of $1.4~\mathrm{\mu s}$ and a Gaussian pulse shape.  Number selective pulses are applied during the second half of a symmetric flux waveform which pulses the coupler to the on-position. Note that the extracted value of $\chi_{sa} / 2\pi = -0.851\pm 0.007~\text{MHz}$ is reported in \cref{table:comprehensive_device_properties} (in the parity measurement position properties section).

\section{Measurement of the off-position storage-ancilla dispersive interaction strength} \label{app:off_position_chi}

\begin{figure}[t!]
    \centering
    \includegraphics[width=\columnwidth]{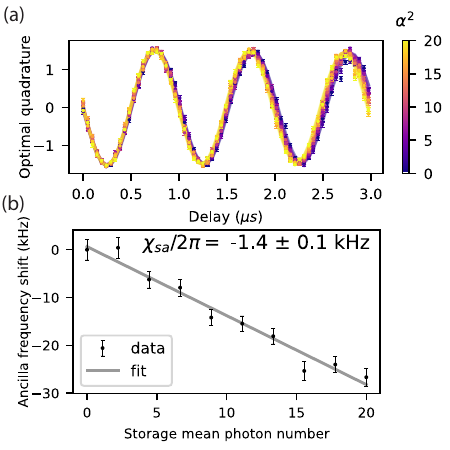}
    \caption{\textbf{Measurement of storage-ancilla $\chi_{sa}$ with coupler in the off-position.} (a) Ramsey measurements of the ancilla for varying storage populations various values of the mean photon number in the storage mode (ranging from $0$ to $20$). (b) Fit of the ancilla frequency as a function of the storage mode mean photon number to determine the storage-ancilla $\chi_{sa}$.}
    \label{app_fig:off_position_chi}
\end{figure} 
The storage conditional phase measurement is not effective at resolving the vanishingly small storage-ancilla dispersive coupling strength in the off-position. Thus for the off-position $\chi_{sa}$ measurement, we instead amplify the effects of the dispersive coupling by driving the storage into a highly excited coherent state (with a mean photon number up to $20$).  Specifically, we perform an ancilla Ramsey experiment with the storage in a variable amplitude coherent state as shown in \cref{app_fig:off_position_chi}(a)  The ancilla experiences an average shift which scales as  $\chi_{sa}|\alpha|^{2}$.  The ancilla frequency shift as a function of the storage photon number is shown in \cref{app_fig:off_position_chi}(b).  A linear fit to the frequency yields a dispersive coupling of $\chi_{sa}/2\pi=-1.4\pm 0.1 \mathrm{kHz}$.

\section{Storage $T_{1}$ and $T_{2}$ measurements}
\label{app:storage_t1_t2_msmts}

\begin{figure}[t!]
    \centering
    \includegraphics[width=\columnwidth]{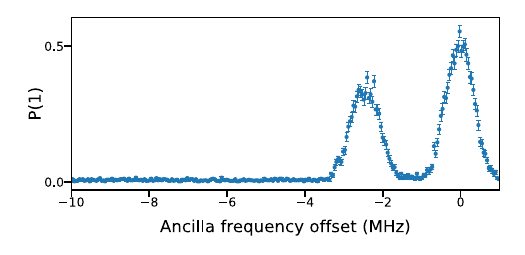}
    \caption{\textbf{Dissipative preparation of state in the $|\hat{n}=0\rangle/|\hat{n}=1\rangle$ manifold.} Ancilla number splitting after applying two-photon dissipation for $8~\mathrm{\mu s}$ starting from $|\alpha|^2=4$.}
    \label{app_fig:dissipative_preparation}
\end{figure}

For both the $T_{1}$ and $T_{2}$ measurements we start off by preparing the storage mode into the $|\hat{n}=0\rangle$/$|\hat{n}=1\rangle$ manifold using a pure two-photon dissipation without the buffer drive~\cite{MarquetAutoparametric2024}.  Specifically we displace the storage mode to $|\alpha|^2=4$ and apply the pure two-photon dissipation for $8~\mathrm{\mu s}$ (number-splitting spectroscopy of the prepared state is shown in \cref{app_fig:dissipative_preparation}). Next we wait for a variable delay time with or without the pure two-photon dissipation being applied.  

For $T_{1}$ measurements the final step is measuring the storage mode parity.  Measurement of the storage mode parity is performed with the coupler pulsed to the parity measurement flux position.  The parity readout is symmetrized by using two phases for the final ancilla $\pi/2$-pulse of the parity measurement~\cite{sun2014}.  

For $T_{2}$ measurements, in the final steps we displace the storage mode and measure the storage mode vacuum population using a weak photon number selective $\pi$-pulse applied to the transmon ancilla.  The storage mode displacement has an amplitude of $\alpha = 0.83 \simeq \ln{\sqrt{1/2}}$, which ensures that when the storage decays to vacuum during the variable time delay, the final displacement results in approximately half of the population remaining in the vacuum state.  In order to ensure that the final measurement is accurate before the steady state is reached, we symmetrize the final displacement.  Specifically, we perform the final measurement with both positive and negative displacements.  Without symmetrization, the use of a frame detuning during the Ramsey sequence results in a nonsymmetric decay envelope.

The reported storage $T_{1}$ ($T_{2}$) values are determined by performing $5$ ($4$) interleaved $T_{1}$ ($T_{2}$) measurements without and with the pure two-photon dissipation being applied during the variable time delay.  All these interleaved data sets for each case are combined together and fit to determine $T_{1}$ and $T_{2}$. The storage $T_{1}$ and $T_{2}$ measurements reported in the main text were taken directly after each other.  On the other hand the measurement of the storage mode phase-flip rate was taken many hours later.  Storage $T_{1}$ measurements with dissipation at the time of the phase-flip rate measurement yielded $74\pm 1~\mathrm{\mu s}$. 

\begin{figure}[t!]
    \centering
    \includegraphics[width=\columnwidth]{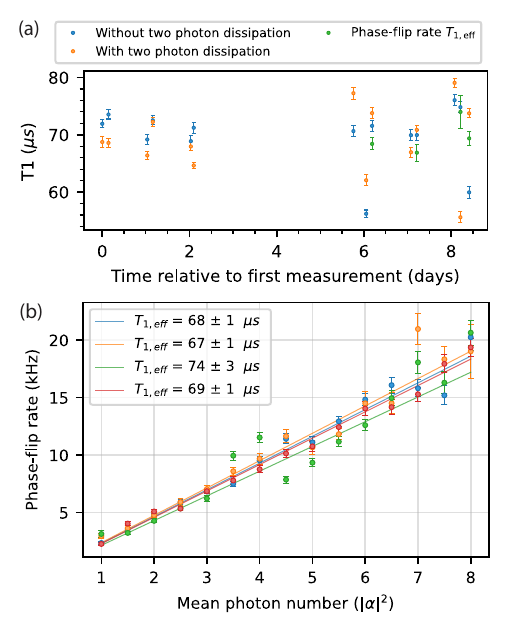}
    \caption{\textbf{Temporal variation of storage $T_{1}$.} (a) Measurements of the storage $T_{1}$ with (orange data points) and without (blue data points) dissipation being applied over time using the same procedure as \cref{fig:storage_coherence}. We also include points representing the $T_{1,\mathrm{eff}}$ (green data points) extracted from fitting the phase-flip rate as a function of $|\alpha|^2$.  (b) Data and fits corresponding to the phase-flip rate experiments that $T_{1,\mathrm{eff}}$ was determined from in (a). Data is averaged over the even and odd initial cat states.}
    \label{app_fig:t1_variation}
\end{figure}

In \cref{app_fig:t1_variation}(a) we show the results of the storage $T_{1}$ measurements intermittently performed over the course of a week.  Each data point uses the same interleaving procedure. Finally we remark that weeks before the data presented here we have observed lower $T_{1}$ ($<50~\mathrm{\mu s}$) on the storage mode. We also include examples of effective lifetimes taken from measurements of phase-flip rates.  The phase-flip rate data corresponding to these points is shown in \cref{app_fig:t1_variation}(b).

\section{Storage mode displacement and amplitude calibrations}
\label{app:displacement_dissipation_amplitude_calibration}
\begin{figure}[t!]
    \centering
    \includegraphics[width=\columnwidth]{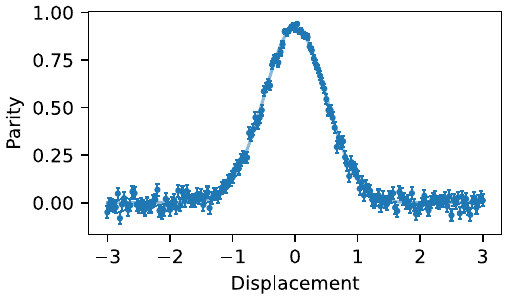}
    \caption{\textbf{Storage mode displacement calibration.} Storage mode displacement is calibrated by fitting the displaced vacuum state parity.  }
    \label{app_fig:displacement_calibration}
\end{figure}

Here we detail the methods we use for calibrating both the storage mode displacement and buffer drive amplitude.  Storage mode displacement is calibrated by fitting the displaced vacuum state parity which is expected to follow $P(\alpha)\propto e^{-2|\alpha|^2}$, where $\alpha$ is the storage mode displacement~\cite{Kirchmair2013}.  An example dataset and fit is shown in \cref{app_fig:displacement_calibration}.  In this experiment we symmetrize the parity measurement readout by using two phases for the final ancilla $\pi/2$ pulse~\cite{sun2014}.

\begin{figure}[t!]
    \centering
    \includegraphics[width=\columnwidth]{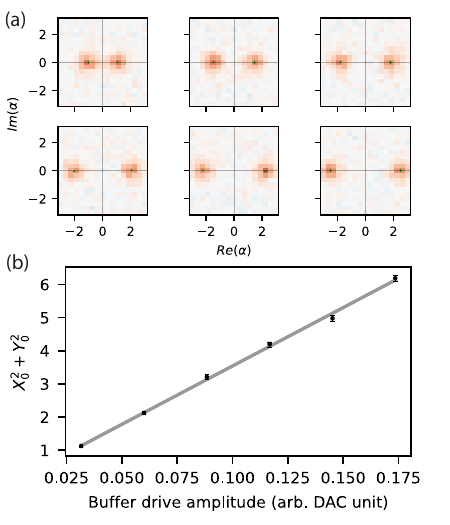}
    \caption{\textbf{Buffer drive amplitude calibration.} (a) Measured Wigner tomogram of the steady state of the storage mode during two-photon dissipation for a series of buffer drive amplitudes. From left to right and top to bottom, the buffer drive amplitude is increasing from $\sim0.03$ to $\sim 0.175$ (arbitrary units). (b) For each Wigner tomogram in (a), corresponding to a buffer drive amplitude, we plot the squared distance from the origin in phase-space ($X^2_{0}+Y^2_{0}$) determined by fitting the Wigner tomogram to two diametrically-opposed Gaussians. We use the linear fit to this curve as a calibration of storage mean photon number versus buffer drive amplitude.}
    \label{app_fig:dissipation_amplitude_calibration}
\end{figure}

To calibrate the relation between the buffer drive amplitude and the size (i.e., average photon number $|\alpha|^{2}$) of the steady-state cat qubit states under the two-photon dissipation, we measure the storage mode Wigner functions for a few different values of the buffer drive amplitude as shown in \cref{app_fig:dissipation_amplitude_calibration}(a)~\cite{Lescanne2020}.  We reach the steady state manifold by applying two-photon dissipation for $200~\mathrm{\mu s}$.  For each buffer drive amplitude we fit the storage mode Wigner function to two diametrically opposed Gaussians.  For points near $|\alpha|^2=1$ this is not exactly the correct model because the storage mode parity is non-zero.  In simulations we find that this effect only causes fits at $|\alpha|^2=1$ to disagree with the storage mode photon number by approximately $3\%$.  We fit the storage mode steady state mean photon number as a function of buffer drive amplitude to a linear model to complete the calibration as shown in \cref{app_fig:dissipation_amplitude_calibration}(b).  

\section{Storage self-Kerr fitting} 
\label{app:storage_kerr}

\begin{figure}[t!]
    \centering
    \includegraphics[width=\columnwidth]{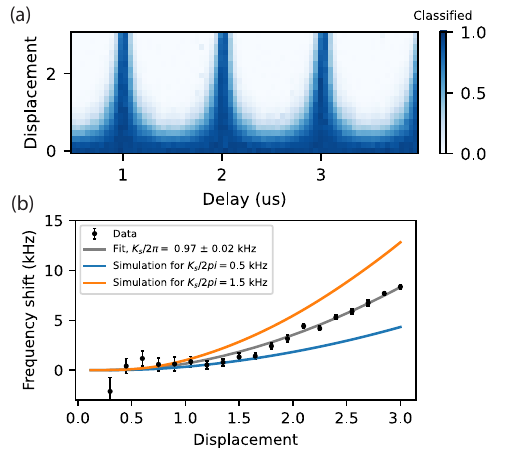}
    \caption{\textbf{Storage self-Kerr fitting.}  (a) Measured occupancy of the storage mode vacuum (color scale bar) versus displacement amplitude and time delay for a storage Ramsey measurement~\cite{Reinhold2019Thesis}.  (b) Plot of the frequency offset of the storage mode frequency versus displacement amplitude of the storage mode, from fits to Ramsey measurements as in (a).  A fit to a simulation of the experiment yields $K_{s}/2\pi= 0.97\pm 0.02~\text{kHz}$. Simulations for $K_{s}/2\pi=0.5$~kHz and $1.5$~kHz are shown as red and blue curves. Note, two points with very low displacements are excluded since they have poor contrast for fitting. Also the frequency offset is defined relative to the average of the first four data points of the curve.}
    \label{app_fig:kerr_fitting}
\end{figure}

To determine the storage mode self-Kerr nonlinearity we perform a storage Ramsey experiment for coherent states of the storage mode with varying coherent-state amplitudes.  This involves displacing the storage mode, waiting for a variable time, displacing the storage mode back with a time dependent phase, and finally measuring the storage mode vacuum population~\cite{Reinhold2019Thesis}.  Note that since the storage mode can be highly excited in this experiment we use a coupler flux position for the number selective measurements near that of the parity measurement flux position.  The vacuum populations from running this experiment are shown in \cref{app_fig:kerr_fitting}(a).  We determine the average storage frequency for each displacement by fitting to the functional form 
\begin{align}
    A e^{-|\alpha|^2 (2-2\cos{(2\pi f t)})}
\end{align}
The free parameters are $A$ which is a scaling factor and $f$ which is the frequency of the storage mode. In \cref{app_fig:kerr_fitting}(b) we show the fit storage mode frequency offset as a function of the displacement from the experimental data.  Fitting the frequency offset curve to a simulation of the experiment yields a storage self-Kerr nonlinearity of $K_{s}/2\pi= 0.97\pm 0.02~\text{kHz}$. 

\section{Two-photon dissipation with ancilla and coupler excitations}
\label{app:dissipation_with_excitations}

\begin{figure}[t!]
    \centering
    \includegraphics[width=\columnwidth]{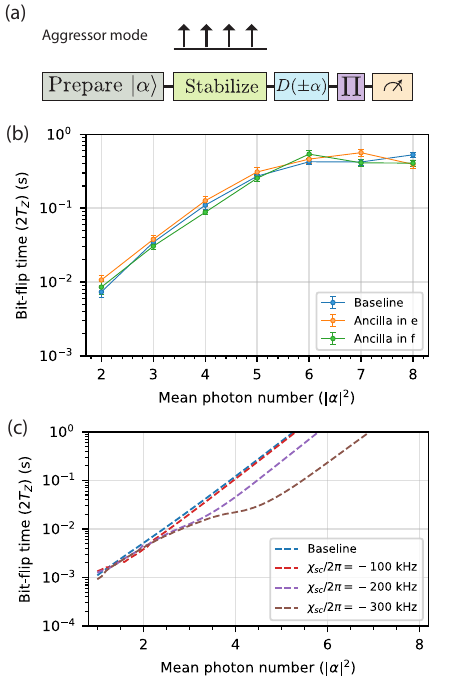}
    \caption{\textbf{Bit-flip times with aggressor excitation.}  (a) Experimental protocol used to perform bit-flip characterization while exciting an aggressor mode.  In the data presented here the aggressor mode is excited every $500~\mathrm{\mu s}$.  (b) Measured bit-flip times when the aggressor mode is the transmon ancilla, from which we infer that excitation of the ancilla does not affect the storage mode bit-flip times. (c) Numerically simulated bit-flip times when a thermally excited coupler mode is included. The different dashed curves (blue, red, purple, brown) correspond to storage-coupler dispersive couplings of $\chi_{sc}/2\pi = (0, -100, -200, -300)~\text{kHz}$, respectively.}
    \label{app_fig:dissipation_aggressor_excitation}
\end{figure} 

To provide further insights into the saturation behavior of the bit-flip times of our cat qubit, we measure the bit-flip times while we repeatedly excite an aggressor qubit (sequence shown schematically in \cref{app_fig:dissipation_aggressor_excitation}(a)).  If the frequency shift from the aggressor qubit excitations are sufficiently strong to overcome the  confinement rate of the two-photon dissipation we expect to observe lower saturated bit-flip times than what we observed without the aggressor qubit excitations. In the sequence we excite the aggressor qubit every $500~\mathrm{\mu s}$, far faster than the observed saturated bit-flip time ($\sim 0.5$s) in the absence of the aggressor excitation.  Note that in the experiment here, the value of the buffer detuning (i.e., the difference between the buffer drive frequency and the Stark-shifted buffer frequency) is slightly different from the one used to generate the data in the main text. However, the saturated bit-flip times are not sensitively dependent on the buffer detuning and we observe similar saturated bit-flip times even with a slightly different buffer detuning in this experiment. 

In \cref{app_fig:dissipation_aggressor_excitation}(b) we show the bit-flip time of our cat qubit when the aggressor qubit is the ancilla.  The blue curve is a reference where the ancilla is not excited.  We consider the case of repeatedly exciting the ancilla to $|e\rangle$ and $|f\rangle$.  In both cases we observe no significant change in the saturated bit-flip times. This is consistent with the small storage-ancilla dispersive shift (i.e., $|\chi_{sa}| = 2\pi \times 1.4~\text{kHz} \ll \kappa_{2}|\alpha|^{2} $) measured with the coupler in the off position (see \cref{app:off_position_chi}).

Since the storage mode is directly coupled to the couplers, the dispersive coupling of the storage mode to the coupler resonances are more sizable (e.g., predicted to be approximately $-100~\text{kHz}$ in the off-position by the model in \cref{app:tunable_dispersive_coupling_details}) than the dispersive coupling to the ancilla.  In \cref{app_fig:dissipation_aggressor_excitation}(c), we numerically study the impact of such storage-coupler dispersive coupling. In particular, in addition to the storage-buffer Hamiltonian and dissipation terms considered in \cref{app:effective_model_analysis}, we further add the storage-coupler dispersive coupling $\chi_{sc}\hat{a}^{\dagger}\hat{a}\hat{c}^{\dagger}\hat{c}$ to the Hamiltonian, where $\hat{c}$ and $\hat{c}^{\dagger}$ are the annihilation and creation operators of the coupler mode. Additionally, we add additional dissipators associated with the coupler loss $\kappa_{c}\mathcal{D}[\hat{c}]$ and heating $\kappa_{c,\text{up}}\mathcal{D}[\hat{c}^{\dagger}]$ to the Lindblad dissipator. Guided by the circuit-quantization predictions, we consider storage-coupler dispersive couplings of $\chi_{sc}/2\pi = 0, -100, -200, -300~\text{kHz}$. Also we assume $\kappa_{c} = 1 / (10~\mathrm{\mu s})$ and $\kappa_{c,\text{up}} / \kappa_{c} = 10^{-3}$.    

The simulation results suggest that the cat-qubit bit-flip times are not degraded when the storage-coupler dispersive coupling strength is $\chi_{sc}/2\pi = -100~\text{kHz}$ (dashed red curve) compared to the baseline case with $\chi_{sc}=0$ (dashed blue curve). Notably these simulations include higher excited states of the coupler up to the third excited state. Thus they account for enhanced dispersive shift of the storage mode frequency under these highly excited coupler states. For larger values of storage-coupler dispersive coupling with $\chi_{sc}/2\pi = -200~\text{kHz}$ (dashed purple curve) and $\chi_{sc}/2\pi = -300~\text{kHz}$ (dashed brown curve), a degradation in the bit-flip time starts to become noticeable, especially in the small mean photon number regime, e.g., $|\alpha|^{2} \lesssim 4$. However, these coupler-induced limitations are overcome, and the exponentially increasing trend of the bit-flip times is quickly recovered at larger values of $|\alpha|^{2} \ge 4$ as the confinement rate of the two-photon dissipation surpasses the storage-coupler dispersive shift. In stark contrast, the cat-qubit bit-flip times in our experiment stay saturated at around $0.3~\text{s}-0.6~\text{s}$ even at large values of $|\alpha|^{2}$ above $10$. These results show that heating of the ancilla and coupler, even to higher excited states beyond the first excited state, are likely not the reason for the saturated bit-flip times of our cat qubit.

\section{Simulation of loop inductance}
\label{app:loop_inductance_simulation}

The inductances $L_{P,1}, L_{P,2}$ in series with the ATS side junctions are estimated during device design from finite-element magnetostatic simulation of the buffer.
In this type of simulation, the ATS metal geometry is modeled as a perfectly conducting surface as though the junctions are electrical shorts, with a junction lead on each side interrupted by a current source.
For each current source, the magnetic field solution is computed by applying current on that source, leaving the other source electrically open.
Each solution thus corresponds to current flowing around one side loop.

The self-inductance of each side loop is calculated from the magnetic energy in the corresponding solution, while a mutual inductance between the two loops is calculated based on the excess energy of the superposition of the two solutions.
We attribute the loop self-inductances mainly to current in the side and middle junction leads, while the mutual inductance is attributed mainly to the shared current path through the middle.
Specifically, we interpret the results as an inductance matrix describing a lumped-element circuit consisting of a loop with inductors $L_{P,1}, L_{P,2}$ on each side, corresponding to the outer SQUID loop geometry of the ATS, which is bisected by an inductance equivalent to the calculated mutual inductance.

To estimate the desired serial inductances, we thus subtract the mutual inductance from each loop's self-inductance. This yields an inductance around 20~pH on each side, compared to 27~pH from the fit to experimental data in \cref{fig:serial_inductance}(b). We note that other effects (e.g., kinetic inductance in the junction leads) which are not included in the above inductance calculations might account for the discrepancy.

\section{Flux-line-induced loss}

The flux lines used to control the buffer also present an environment to the buffer without the protection of a bandpass filter. For that reason, photon loss through the flux lines may still affect the storage mode's lifetime, even if the buffer mode's lifetime is dominated by loss through its filtered output line.

To understand loss induced by the flux lines, we consider an isolated buffer as in \cref{app:pump_loss}, statically parked at a saddle point, but now allowing for time-dependent flux due to fluctuating current in both buffer flux-pump lines. Following the procedure for circuit quantization with time-dependent flux described in Refs.~\cite{You2019_circuit,Riwar2021_circuit}, we obtain a Hamiltonian in terms of ``irrotational'' degrees of freedom $\tilde{N}_b,\tilde{\phi}_b$ such that all time-dependent pump fluctuations are assigned to the inductive potential:

\begin{align}
    H_\mathrm{irr}(\tilde{N}_b,\tilde{\phi}_b) &= 4E_{C} \tilde{N}_b^2 \notag\\
    &-  N E_{J,\mathrm{array}} \cos\frac{1}{N}\Big{(}\tilde{\phi}_b - \varphi_\Delta(t)  + \varphi_\delta(t)\Big{)} \notag\\
    &- (E_{J,1} + E_{J,2}) \sin \varphi_\Sigma(t) \sin\left(\tilde{\phi}_b + \varphi_\delta(t)\right) \notag\\
    &- (E_{J,1} - E_{J,2}) \cos \varphi_\Sigma(t) \cos\left(\tilde{\phi}_b + \varphi_\delta(t)\right).
\end{align}

Above, the sigma-flux fluctuation $\varphi_\Sigma(t)$ is assigned to the side junctions, while the delta-flux fluctuation $\varphi_\Delta(t)$ is assigned to the array.
Each branch also sees a common offset $\varphi_\delta(t)$, which can be interpreted as related to magnetic field nonuniformity in the buffer's capacitive gap (not just the ATS loops).
We describe these fluxes as arising from inductive couplings to current $I_\Sigma(t), I_\Delta(t)$ in each flux line.
To estimate rates of photon emission into the flux lines, we then expand the above Hamiltonian assuming small fluctuations and consider linear couplings $I(t) \tilde{\phi}_b$.

For example, we consider the case where the only time-dependent flux is $\varphi_\Sigma(t) = \frac{2\pi}{\Phi_0} M_\Sigma I_\Sigma(t)$, where the inductive coupling  $M_\Sigma$ is required for operating the buffer pump. That is, we ignore both $\varphi_\Delta(t)$ and $\varphi_\delta(t)$, as well as any unintended coupling between the delta-flux line and the sigma flux. This yields the term of interest $- (E_{J,1} + E_{J,2}) \frac{2\pi}{\Phi_0} M_\Sigma I_\Sigma(t) \tilde{\phi}_b$, corresponding to a loss rate
\begin{align}
    \kappa_\Sigma = \frac{\phi_\mathrm{zpf}^2}{\hbar^2} \left( \frac{2\pi}{\Phi_0}  (E_{J,1} + E_{J,2}) M_\Sigma \right)^2 S_{I_\Sigma I_\Sigma}(\omega)
\end{align}
for a mode at frequency $\omega$ with zero-point fluctuation $\phi_\mathrm{zpf}$ associated with $\phi_b$, where $S_{I_\Sigma I_\Sigma}(\omega)$ is the noise power spectrum of the current $I_\Sigma(t)$. For example, if the buffer sees a real load impedance of $Z_\Sigma$ at zero temperature through the flux line, then $S_{I_\Sigma I_\Sigma}(\omega) = 2 \hbar \omega/Z_\Sigma$. Together with finite element simulation to estimate inductive couplings, further calculation shows that this is the largest expected contribution to flux-line-induced loss. Accordingly, our flux line design targets a minimal $M_\Sigma$.

\bibliographystyle{naturemag}
\bibliography{single_cat} 

\end{document}